\documentclass[fleqn,usenatbib]{mnras}

\usepackage[T1]{fontenc}
\usepackage{soul}

\DeclareRobustCommand{\VAN}[3]{#2}
\let\VANthebibliography\thebibliography
\def\thebibliography{\DeclareRobustCommand{\VAN}[3]{##3}\VANthebibliography}

\usepackage{graphicx}	
\usepackage{amsmath}	
\usepackage{amssymb}	

\title[Truncated accretion discs]{Truncated accretion discs in black hole X-ray binaries: dynamics and variability signatures}

\author[Dihingia et al.]{
Indu K. Dihingia,$^{1,2,3}$\thanks{E-mail: idihingia@iiti.ac.in}
Bhargav Vaidya,$^{1}$
and Christian Fendt$^{4}$
\\
$^{1}$Department of Astronomy, Astrophysics and Space Engineering, Indian Institute of Technology Indore, Khandwa Road, Simrol, 453552, India\\
$^{2}$Department of Physics, Indian Institute of Science, Bangalore 560012, Karnataka, India\\
$^{3}$Tsung-Dao Lee Institute, Shanghai Jiao-Tong University, Shanghai, 520 Shengrong Road, 201210, People's Republic of China\\
$^{4}$Max Planck Institute for Astronomy, K\"onigstuhl 17, 69117 Heidelberg, Germany\\
}

\date{Accepted XXX. Received YYY; in original form ZZZ}

\pubyear{2015}

\begin{document}
\label{firstpage}
\pagerange{\pageref{firstpage}--\pageref{lastpage}}
\maketitle
\newcommand{\bvc}[1]{\textbf{\color{green}#1}}

\begin{abstract}
Variable features in black hole X-ray Binaries (BH-XRBs) are observed in different energy ranges and time scales. The physical origin of different spectral states in BH-XRBs and their relations with the underlying accretion disc are still elusive. To investigate the intermediate state of BH-XRBs during outburst, we simulate a truncated accretion disc around a Kerr black hole using a general relativistic magneto-hydrodynamical (GRMHD) framework under axisymmetry with adaptively refined mesh. Additionally, we have also carried out radiative transfer calculations for understanding the implications of disc dynamics on emission. Dynamically, the inner edge of the truncated accretion disc oscillates in a quasi-periodic fashion (QPO). The QPO frequency of oscillations $(\nu_{\rm QPO, max})$ increases as the magnetic field strength and magnetic resistivity increase.  However, as the truncation radius increases, $\nu_{\rm QPO, max}$ decreases. In our simulation models, frequency varies between $7\times(10M_{\odot}/M_{\rm BH})$ Hz $\lesssim\nu_{\rm QPO, max}\lesssim20 \times (10M_{\odot}/M_{\rm BH})$ Hz, which is in the range of low-frequency QPOs.
We further find evidence of transient shocks in the highly accreting stage during oscillation. Such a transient shock acts as an extended hot post-shock corona around the black hole that has an impact on its radiative properties. The radiative transfer calculations show signatures of these oscillations in the form of modulation in the edge-brightened structure of the accretion disc.
\end{abstract}




\section{Introduction}
X-ray binaries (XRBs) and active galactic nuclei (AGNs) typically have spectral and temporal variations in their light curves, which could be the imprints of physical processes occurring around the central objects. The BH-XRBs are mostly inactive, with rare outbursts caused by a sudden surge in accretion activity. In BH-XRBs, an outburst and its various spectral states (hard state, hard/soft intermediate state, and soft state) characteristics are commonly characterized using a hardness-intensity `Q' diagram \citep[see for detail,][etc.]{Fender-etal2004, Remillard-McClintock2006,Done-etal2007, Dunn-etal2010,Belloni2010, Belloni-Motta2016}. 

The thermal component dominates the spectrum in soft-state, and it may be understood using a geometrically thin, optically thick accretion disc that produces a blackbody spectrum locally \citep {Shakura-Sunyaev1973, Novikov-Thorne1973}. In the hard state, a high-energy power-law component dominates the spectrum, and the contribution of both hard and soft components can be seen in the intermediate state \citep[e.g.,][]{Homan-Belloni2005, McClintock-etal2006, Belloni2010}. Compton upscattering of soft X-ray photons by hot electrons present in the corona close to the black hole causes the power-law component \citep{Thorne-Price1975, Sunyaev-Truemper1979, Chakrabarti-Titarchuk1995}.
The corona also plays an important role in understanding the observed correlation between different energy ranges of BH-XRBs \citep[e.g.,][]{Kylafis-etal2018}. Recently, with a correlation study for GRS 1915+105, it has been shown that the matter in the corona even contributes to the jet at a different time \citep{Mendez-etal2022}. Thus, it is crucial to study corona formation close to the black hole and investigate its dynamics for understanding the observed variability in BH-XRBs. There are essentially two models with regard to the placement of the corona: (i) lamp-post model, and (ii) extended corona \citep[see][]{Chauvin-etal2018}. The lamp-post model posits a compact corona at a specific height along the black hole's angular momentum axis. This is a simplified model and is often considered as ad-hoc \citep[and references therein] {Degenaar-etal2018}.

On the other hand, the extended corona model has several flavors depending on where the corona is located \citep{Galeev-etal1979, Haardt-Maraschi1991, Miyamoto-Kitamoto1991, Fender-etal1999, Kylafis-Belloni2015}. However, the general agreement is that it originates from the thin disc itself and occupies an area near the black hole. Evaporation of a truncated accretion disc at a certain radius can produce such a region \citep{Eardley-etal1975, Ichimaru1977}. Also, because of the shock-transition in the accretion flow, such corona can be produced from low angular momentum flow \citep[references therein]{Das2007,Chakrabarti-etal2015,Dihingia-etal2018,Dihingia-etal2019a,Dihingia-etal2019b,Dihingia-etal2020}. The generation of shock waves and oscillations is demonstrated using asymmetric hydrodynamical simulations \citep{Lee-etal2011,Lee-etal2016,Das-etal2014}. \cite{Singh-etal2021} recently have shown the production of shock with resistive MHD. However, the shock formation in these simulations is highly dependent on the initial conditions, and it requires supersonic injection of accretion materials at the outer boundary.

General relativistic magneto-hydrodynamics (GRMHD) is an indispensable technique for studying the physics of active galactic nuclei (AGNs) \citep[see,][references therein]{Davis-Tchekhovskoy2020, Mizuno2022}. However, GRMHD has not been widely used to investigate the physics of BH-XRBs.  Recently, \cite{Dexter-etal2021} used radiation GRMHD around BH-XRBs and attributed the strongly magnetized accretion to the hard-state. In our recent study \citep{Dihingia-etal2021,Dihingia-Vaidya2022a}, we simulated a high-angular momentum thin disc around the black hole. We initially observed a structured jet and disc-wind around the black hole, but after sufficient simulation time, the inner part of the accretion disc starts to oscillate, and the time period of oscillation increases with time. We attributed the oscillating phase to be the hard-intermediate state (HIMS) seen in BH-XRBs.

The onset of an outburst happens once a thin-disc (high angular momentum matter) forms at a large distance $(\sim2\times10^4$ gravitational radius ($r_g$)) from the black hole \citep[see for discussion,][]{Kylafis-Belloni2015}. Such a scenario can be realized as a truncated accretion disc with a very large truncation radius around a black hole. Simulating such a realistic picture is computationally expensive. Consequently, we chose a truncation radius closer to the black hole to capture physics only in an intermediate state of the initial phase of the outburst. 
Following this, we simulate a truncated accretion disc around a Kerr black hole. We consider that the standard thin disc is only up to a certain radius $(r_{\rm tr})$. Due to this generality, the truncated accretion disc may help us in understanding the physically motivated corona model and its variability signatures. In this study, we investigate the properties of the truncated accretion disc and its behavior with different truncation radii and magnetic field strengths both for the ideal MHD case and with different resistivity for the resistive case. Further, we employ the general relativistic radiative transfer (GRRT) post-processing module RAPTOR \citep{Bronzwaer-etal2018} to understand its consequences in the radiative properties of the truncated accretion disc. 

Our paper is arranged as follows: In section 2, we explain in detail the numerical approach along with the initial conditions considered for our study. Section 3 explains the dynamical evolution of simulation models. Section 4 provides temporal characteristics of time average quantities, and in section 5, we present the radiative properties of simulation models. A summary and discussion of our results are presented in section 6.
\section{Numerical setup}
\subsection{Basic equations}
For this work, we use the \texttt{code BHAC} equipped with adaptive mesh refinement (AMR) to solve the set of ideal as well as resistive GRMHD equations following \cite{Porth-etal2017, Olivares-etal2019,Ripperda-etal2019}. The basic resistive GRMHD equations are as follows, 
\begin{align}
\begin{aligned}
&\nabla_\mu\left(\rho u^\mu\right)=0,\\
&\nabla_\mu T^{\mu\nu}=0,\\
&\nabla_\mu F^{\mu\nu}={\cal J}^\mu,\\
&\nabla_\mu{}^*F^{\mu\nu}=0,\\
\end{aligned}
\label{eq-01}
\end{align}
where different symbols have their usual meaning, viz. $\rho, u^\mu, T^{\mu\nu}, F^{\mu\nu}, {}^*F^{\mu\nu}$, and ${\cal J}^{\mu}$, stand for rest-mass density, four velocity, energy-momentum tensor, Faraday tensor, dual of the Faraday tensor, and electric 4-current, respectively. 
The details of these terms and explicit procedure of solving is given in \cite{Porth-etal2017,Olivares-etal2019,Ripperda-etal2019}. These equations are solved in a spherically symmetric, Modified Kerr-Schild (MKS) geometry in axisymmetric. 
By choosing an appropriate MKS stretching parameter, we ensure that the concentration of maximum resolution is near the equatorial plane of the simulation domain \citep{McKinney-Gammie2004}. 
\texttt{Code BHAC} employs the constrained-transport method \citep{DelZanna-etal2007} to ensure a divergence-free magnetic field in the simulation domain. Throughout the work, we use a generalised unit system considering  $G=M=c=1$, where $M$, $G$, and $c$ are the mass of the black hole, the universal gravitational constant, and the speed of light, respectively. In this unit system, mass, length, and time are expressed in terms of $M$, $r_g=GM/c^2$, and $t_g=GM/c^3$, respectively. Subsequently, we follow convention of sign of the metric as $(-, +, +, +)$, with four velocities satisfying $u_\mu u^\mu = -1$. 

Our simulation box is extended radially from $r=r_{\rm H}$ (event horizon) to the $r_{\rm out}=500$. In the polar direction, our box is extended from $\theta=0-\pi$.

\subsection{Initial conditions}
We set up a initial thin-disc following \cite{Dihingia-etal2021,Dihingia-Vaidya2022a}, the setup is based on the \cite{Novikov-Thorne1973} model. The density profile in for the thin-disc setup in Boyer-Lindquist coordinates is given by,

\begin{align}
\rho(r,\theta) = \rho_e(r) \exp\left(-\frac{\alpha^2 z^2}{H^2}\right); ~~ z=r\cos(\theta).
\label{eq-rho}
\end{align}
Here, we choose $\alpha=2$ to maintain a thin disc configuration and $H$ is the scale height of the accretion disc, which is obtained following \citet{Riffert-Herold1995} and \citet{Peitz-Appl1997}. The density profile on the equatorial plane $(\rho_e(r))$ is given by,

\begin{align}
\rho_e(x)=\left(\frac{\Theta_0}{\cal K}\right)^{1/(\Gamma -1)} 
              \left(\frac{f(x)}{x^2}\right)^{1/(4(\Gamma - 1))},
\label{eq-rhoe} 
\end{align}
where $x=\sqrt{r}$, $\Theta_0$ is a constant related to the temperature of the initial disc, we chose $\Theta_0 = 0.0001$. Also, we consider polytropic index $\Gamma = 4/3$ and entropy constant ${\cal K}=0.1$ for this study. Finally, $f(x)$ is given by,
\begin{align}
\begin{aligned}
f(x) =& \frac{3}{2} \frac{1}{ x^2 \left(2 a+x^3-3 x\right)}\bigg[ x - x_0 -\frac{3}{2}\ln\left(\frac{x}{x_0}\right) \\
&- \frac{3\left(s_1-a\right)^2}{s_1(s_1-s_2)(s_1-s_3)} \ln \left(\frac{x- s_1}{x_0-s_1}\right) \\
&- \frac{3\left(s_2-a\right)^2}{s_2(s_2-s_1)(s_2-s_3)}\ln \left(\frac{x- s_2}{x_0-s_2}\right) \\
&- \frac{3\left(s_3-a\right)^2}{s_3(s_3-s_1)(s_3-s_2)}\ln \left(\frac{x- s_3}{x_0-s_3}\right)\bigg], \\
\end{aligned}
\label{eq-fx}
\end{align}
where $x_0=\sqrt{r_{\rm ISCO}}$, $r_{\rm ISCO}$ is the radius of the ISCO (innermost stable circular orbit), and $s_1, s_2,$ and $s_3$ are the roots of the equation $s^3 - 3s + 2a=0$. Along with the density profile, we also supply the initial azimuthal velocity, which is given as follows \citep{Dihingia-etal2021},
\begin{align}
u^\phi(r,\theta) = \left(\frac{\cal A}{{\cal B}+ 2 {\cal C}^{1/2}}\right)^{1/2},
\label{eq-12}
\end{align}
where
$$
\begin{aligned}
{\cal A}=&\left(\Gamma^r_{tt}\right)^2,\\
{\cal B}=&g_{tt}\left(\Gamma^r_{tt}\Gamma^r_{\phi \phi}-2 {\Gamma^r_{t\phi}}^2\right)+2 g_{t\phi} \Gamma^r_{tt} \Gamma^r_{t\phi} - g_{\phi \phi } {\Gamma^r_{tt}}^2,\\
{\cal C}=&\left({\Gamma^r_{t\phi}}^2 - \Gamma^r_{tt} \Gamma^r_{\phi \phi}\right) (g_{t\phi} \Gamma^r_{tt}- g_{tt} \Gamma^r_{t\phi})^2.\\
\end{aligned}
$$
Here, $\Gamma^\alpha_{\beta\gamma}$ and $g_{\mu\nu}$ are the non-zero components of the Christoffel symbols and metric for Kerr black hole, respectively. Furthermore, we consider the thin-disc is truncated at a radial distance $r_{\rm tr}$ and extended up to the outer boundary of the simulation box $r_{\rm out}=500$.

\begin{figure}
\centering
\includegraphics[scale=0.42]{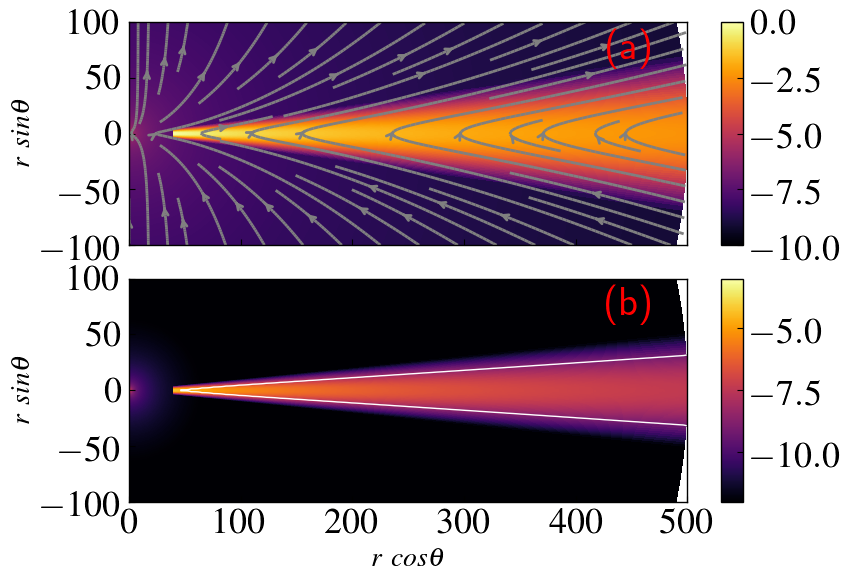}
\caption{Initial (a) logarithmic density ($\log(\rho/\rho^{\rm tr})$) and the (b) logarithmic gas pressure ($\log(p_{\rm gas}/p_{\rm gas}^{\rm tr})$) distribution for the reference model. The gray lines in the upper panel (a) show the initial magnetic field lines, and the white line in the lower panel (b) shows the boundary of plasma-$\beta=1$.}
\label{fig-initial}
\end{figure}

In our study, we consider that the accretion disc is threaded by the poloidal magnetic field lines. The initial poloidal field lines are prescribed by implementing a  vector potential $A_\phi$ following \cite{Zanni-etal2007,Vourellis-etal2019}. The functional form of the vector potential is given by
\begin{align}
A_\phi \propto \left(r \sin \theta\right)^{3/4} \frac{m^{5/4}}{\left(m^2 + \tan^{-2}(\theta-\pi/2)\right)^{5/8}}.
\label{eq-02}
\end{align}
The parameter $m (=0.1$) is related to the initial inclination of the field lines and it also determines the magnetic flux of the system. The parameter $m$ play crucial role in the launching of Blandford-Payne type wind from the accretion disc \citep{Blandford-Payne1982,Dihingia-etal2021}. The initial strength of the poloidal magnetic field is determined by the choice of the plasma-$\beta$ parameter at the truncation radius $r_{\rm tr}$ on the equatorial plane as, $\beta_{\rm tr} = p_{\rm gas}^{\rm tr}/p_{\rm mag}^{\rm tr}$. Here, superscript `tr' denotes quantities calculated at the $r=r_{\rm tr}$.
Following our motivation, we carry out eight axisymmetric simulation models, by choosing different truncation radius $(r_{\rm tr})$, initial plasma-$\beta$ ($\beta_{\rm tr}$), effective resolution (with AMR), and magnetic resistivity. 
For axisymmetric models, we consider the highest effective resolution to be $2048\times1024$ (with three refinement levels). 
The list of input parameters, effective resolution, and the final simulation time $(t_{\rm final})$ for all the simulation models are shown in the table \ref{tab-01}. Out of all these models, we consider 2D40AH to be our reference model for the sake of explanation and comparison. In Fig. \ref{fig-initial}, we show the initial density distribution $(\log(\rho/\rho^{\rm tr}))$ and the initial gas pressure distribution $(\log(p_{\rm gas}/p_{\rm gas}^{\rm tr}))$ for the reference model at panels (a) and (b), respectively. In panel Fig. \ref{fig-initial}a, we also show the initial magnetic field lines in terms of gray lines. In Fig. \ref{fig-initial}b, the white line represent the boundary of plasma-$\beta=1$. In the figure, the density distribution follows equation \ref{eq-rho} outside the truncation radius ($r_{\rm tr}=40$). Near the equatorial plane, the matter distribution is gas pressure dominated, and far from the equatorial plane and inside the truncation radius, the matter distribution is magnetic pressure dominated.

\begin{table}
\centering
  \begin{tabular}{| l | c |c | c | c | c }
    \hline
    Model & $\eta$ &Effective resolution & $r_{\rm tr}$ & $\beta_{\rm tr}$ & $t_{\rm final}$\\ 
    \hline
    2D40A & 0 & $1024\times 512 $ & 40 & 100 & 14000\\
    2D40AH & 0 &$2048\times 1024 $ & 40 & 100 & 9790 \\
    rl-2D40AH & $1\times 10^{-3}$ &$2048\times 1024 $ & 40 & 100 & 9560 \\
    rh-2D40AH & $5\times 10^{-3}$ &$2048\times 1024 $ & 40 & 100 & 8600 \\
    2D40B & 0 & $1024\times 512 $ & 40 & 500 & 14000\\
    2D40C & 0 &$1024\times 512 $ & 40 & 1000 & 12000\\
    2D50 & 0 &$1024\times 512$ & 50 & 100 & 11160\\
    2D60 & 0 &$1024\times 512$ & 60 & 100 & 11480\\
    \hline
  \end{tabular}
\caption{The explicit values of effective resolution, resistivity ($\eta$), truncation radius $r_{\rm tr}$, initial plasma-$\beta$ at $r_{\rm tr}$ ($\beta_{\rm tr}$), and the final simulation time $t_{\rm final}$ for different simulation models.}
\label{tab-01}
\end{table}
\subsection{Boundary conditions}
To prevent material from entering the numerical domain and interfering with accretion flow from the truncated disc, we impose no-inflow conditions at the inner radial boundary. At the polar axis, the scalar and radial components of vectors are considered symmetric, whereas the azimuthal and polar components of vectors are considered antisymmetric.
Furthermore, in our simulation setup, we do not allow any inflow of matter at the outer edge of the accretion disc, which is rather a naive approximation. However, as the current study only focuses to understand the dynamics of the inner part of the truncated accretion disc, we run our simulation models up to $\sim 450-500$ inner disc orbits (at ISCO). This is about $\sim 15\%-20\%$ of the rotation time of the outer edge at $r=500$ and therefore the idealistic far out radial boundary will not affect the inner flow structure significantly within our simulation time. 

\section{Dynamical Characteristics}
\begin{figure*}
\centering
\includegraphics[scale=0.45]{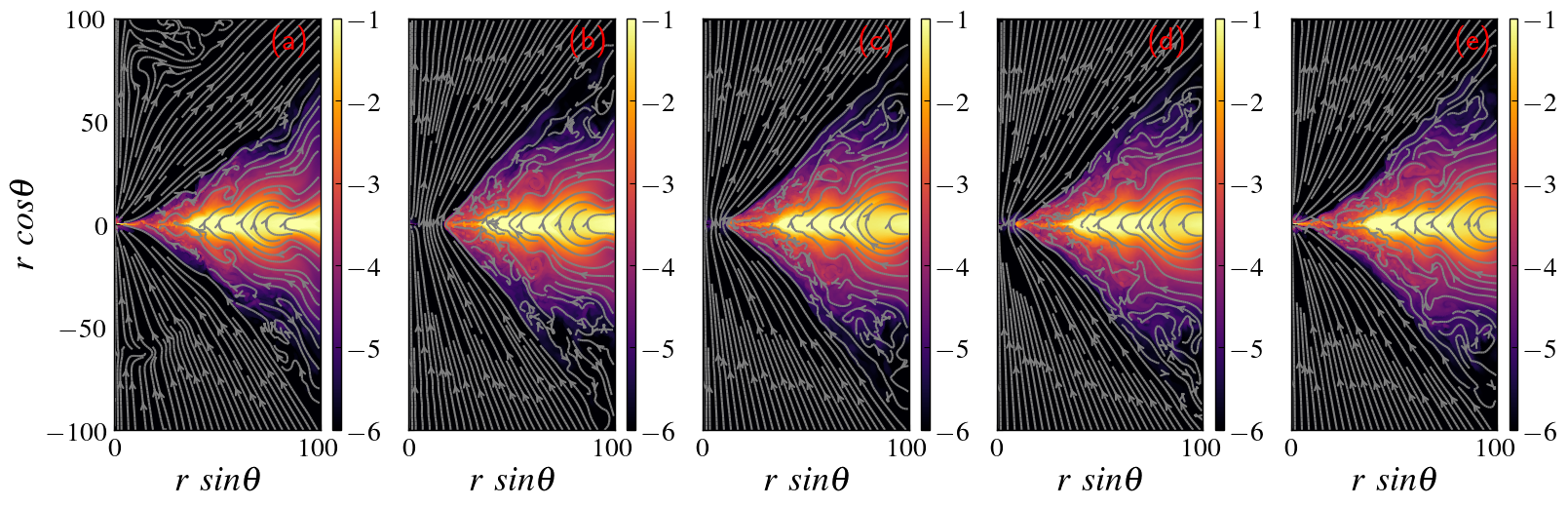}
\caption{Distribution of logarithmic normalised density distribution $(\rho/\rho^{\rm tr})$ and magnetic field lines for reference model 2D40AH in ideal MHD at times (a) $t=3240$, (b) $t=3800$, (c) $t=4080$, (d) $t=4220$, and (e) $t=4380$.}
\label{fig-rho}
\end{figure*}
\subsection{A reference case - ideal MHD}
We first discuss in detail our reference simulation 2D40AH, which is an axisymmetric run involving an ideal MHD approximation. Subsequently, in the later sections, a comparison is done with results obtained from resistive simulations.

The truncated accretion disc winds up the initial weak poloidal magnetic field lines and generates the toroidal component with the temporal evolution. 
The differential rotation of the flow triggers magneto-rotational instabilities (MRI) and drives turbulence, which helps in the transport of angular momentum. To ascertain that MRI is active in our simulations, we set the resolution such that the MRI quality factor $Q_{\theta}\gtrsim10$ (see Appendix A for detail) throughout the simulation domain \citep[see][]{Sano-etal2004}. 
With the angular momentum transport, the Keplerian matter of the inner edge loses angular momentum and becomes sub-Keplerian. Subsequently, sub-Keplerian matter can occupy orbits of lower radii. As a result, the flow starts to accreting towards the black hole. Furthermore, the onset of disc-winds aids in transporting angular momentum outwards, which also contributes to setting up the accretion process \citep{Dihingia-etal2021}.
Eventually, the matter reaches the event horizon of the black hole.

\begin{figure}
\centering
\includegraphics[scale=0.43]{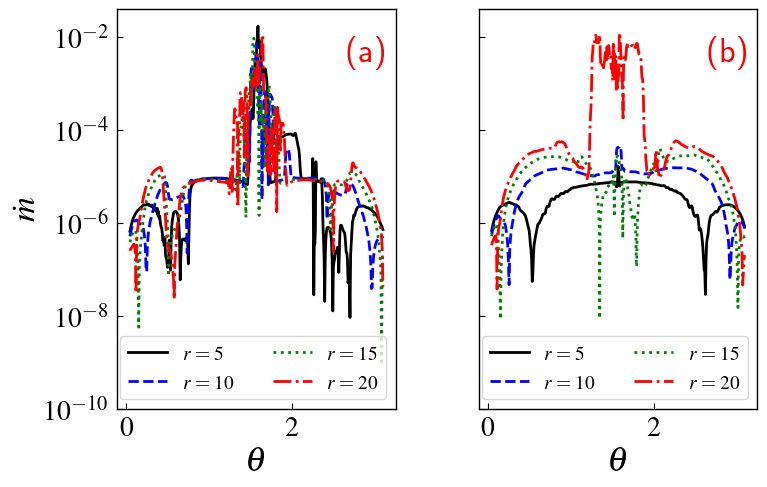}
\caption{Plot of mass flux $\dot{m}=|\sqrt{-g}\rho u^r|$ at different radius labeled on the figure for the reference model at time $t=3240$ (a) and $t=3800$ (b).}
\label{fig-mdot_line}
\end{figure}

To understand the dynamical properties in detail, in Fig. \ref{fig-rho}, we show the logarithmic normalized density distribution $(\rho/\rho^{\rm tr})$ and magnetic field lines for model 2D40AH. The panels (a), (b), (c), and (d) correspond to simulation time $t=3240, 3800, 4080$, and $4380$, respectively. It is interesting to note that in Fig. \ref{fig-rho}a and Fig. \ref{fig-rho}e, the accretion disc is extended up to the event horizon. In between (Fig. \ref{fig-rho}b-\ref{fig-rho}d) the accretion disc is truncated. It essentially suggests that the accretion disc connects to the event horizon and disconnects from the event horizon periodically. During this process, the inner edge of the accretion disc oscillates in a quasi-periodic manner.
We observe such oscillation events throughout our simulation (discussed further in section 4). In Fig. \ref{fig-mdot_line}, we show the mass flux $\dot{m}=|\sqrt{-g}\rho u^r|$  at radii $r=5, 10, 15,$ and $20$ as a function of $\theta$ at simulation time (a) $t=3240$ and (b) $t=3800$ for the reference model. In Fig. \ref{fig-mdot_line}a, we observe very high mass flux around the equatorial plane ($\theta\sim\pi/2$). As a result, matter advects to the black hole efficiently. In this state, matter drags the magnetic field lines along with it, and due to the high mass flux, magnetic flux efficiently accumulates around the event horizon. Further, we also find that these magnetic field lines are rooted into the ergosphere (see Fig. \ref{fig-rho}a and Fig. \ref{fig-rho}e). 
As the magnetic flux accumulated around the horizon reaches $\dot{\phi}_{\rm BH}\gtrsim 15$ (see section 4 for more detail), flow achieves a magnetically arrested disc (MAD) configuration \citep{Tchekhovskoy-etal2011}. However, magnetic flux does not keep on building around the black hole and eventually loses the saturated magnetic flux in a rapid phenomenon commonly known as a magnetic eruption event \citep[e.g.,][]{Igumenshchev2008,Vourellis-etal2019, Porth-etal2021, Dexter-etal2020}. After the eruption event, the accretion disc becomes truncated and the mass flux around the equatorial plane becomes negligible (see the black solid, blue dashed, and green dotted lines in Fig. \ref{fig-mdot_line}b).

\begin{figure*}
\centering
\includegraphics[scale=0.6]{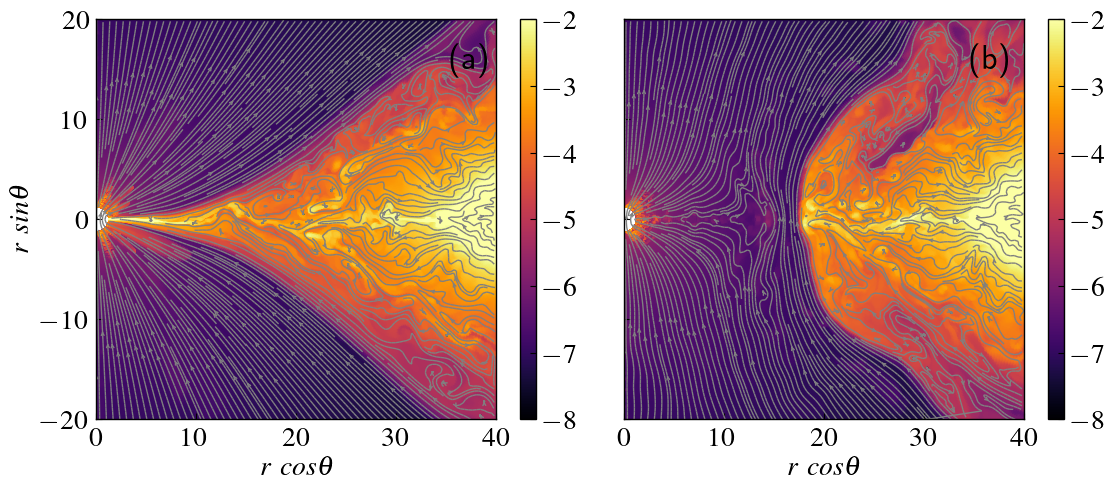}
\caption{Logarithmic density profile and the field lines are shown on the poloidal plane for reference model at time $t=3240$ (a) and $t=3800$ (b).}
\label{fig-zoom}
\end{figure*}

\begin{figure}
\centering
\includegraphics[scale=0.46]{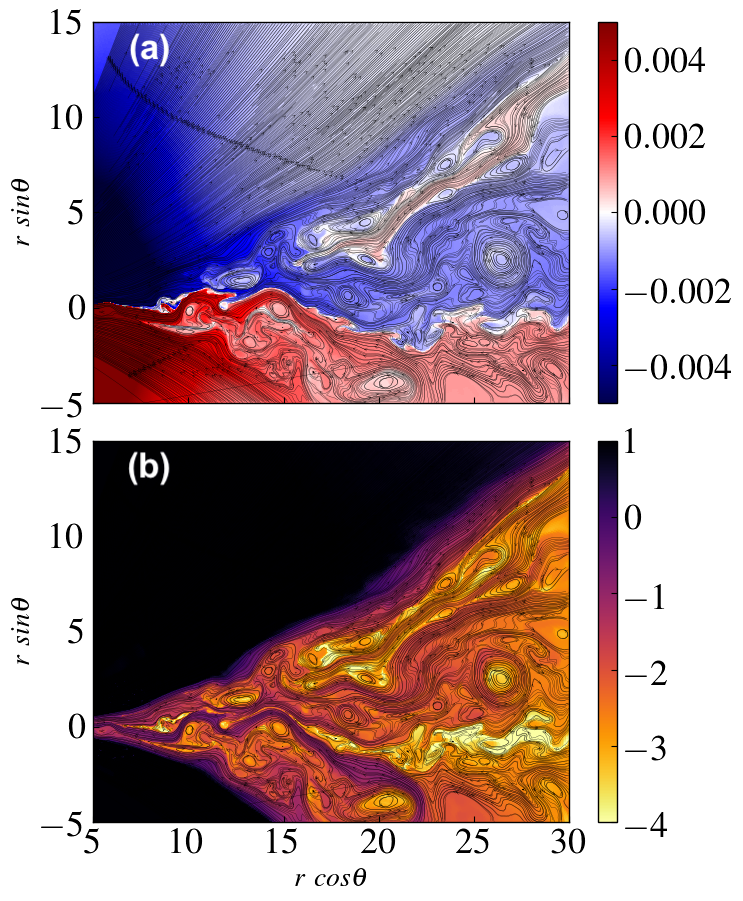}
\caption{Distribution of toroidal component of the magnetic field $B_\phi$ and logarithmic magnetisation $(\sigma=b^2/\rho)$ for the reference model at time $t=4380$. The black solid lines represent the magnetic field lines .
}
\label{fig-bphi}
\end{figure}

To understand the process of oscillation in detail, we show the density profile along with the magnetic field lines at time $t=3240$ (a) and $t=3800$ (b) for the reference model in Fig. \ref{fig-zoom}. In the highly accreting state ($t=3240$), opposite polarity magnetic field lines squeezed together near the black hole, resulting in the formation of current sheets. These current sheets are prone to reconnection events. Furthermore, the turbulent $B_{\phi}$ component changes polarity and facilitates reconnection. In the ideal MHD models, the reconnection events are governed by the grid-dependent numerical resistivity. The distribution of the toroidal component of the magnetic field (a) $B_{\phi}$ and the magnetisation $(\sigma=B^2/\rho)$, as well as the magnetic field lines close to the black hole, are shown in Figure~\ref{fig-bphi} for the reference model at time $t=4380$. The diagram clearly depicts the formation of islands of opposite polarity magnetic fields (\ref{fig-bphi}a), in which low-magnetised matter is trapped within highly magnetised matter (reffig-bphib). In our simulations, these figures (\ref{fig-bphi}a,b) confirm the formation of plasmoids and chains of plasmoids. Our simulations' chain of plasmoids strongly suggests the presence of active tearing mode instabilities.

These reconnection events also create new poloidal field lines penetrating the equatorial (see Fig. \ref{fig-zoom}b). That contributes to a strong magnetic tension force opposing the accretion of the flow. As a result, accretion is halted, and the accretion rate drops by orders of magnitude (for example, see solid black line in Fig. \ref{fig-mdot_line}a and \ref{fig-mdot_line}b).
The increased magnetic tension force momentarily gets balanced with the force due to the ram pressure of the flow at a certain radius $r_{\rm max}$. 
Consequently, for the reference model, the inner edge of the accretion disc recedes up to a radius $r_{\rm max}\sim20$, 
which is two times less than the initial truncation radius $r_{\rm tr}=40$. 
Further, the matter at the inner edge again accretes towards the central black hole due to loss of angular momentum and gravitational attraction forming an highly accreting flow (see the inner edge of the disc in Fig. \ref{fig-rho}b-\ref{fig-rho}e). We observe the formation and depletion of highly accreting flow repeated till the end of our simulation. As a result, the inner part of the accretion disc keeps on oscillating between a highly accreting state and a low accreting state.
Note that the radius $r_{\rm max}$ does not remain the same throughout the simulation time, $r_{\rm max}$ decreases with simulation time. For example, the edge of the accretion disc for the reference model recedes up to $r_{\rm max}\sim17$ at simulation time $t\sim10000$.

\begin{figure}
\centering
\includegraphics[scale=0.40]{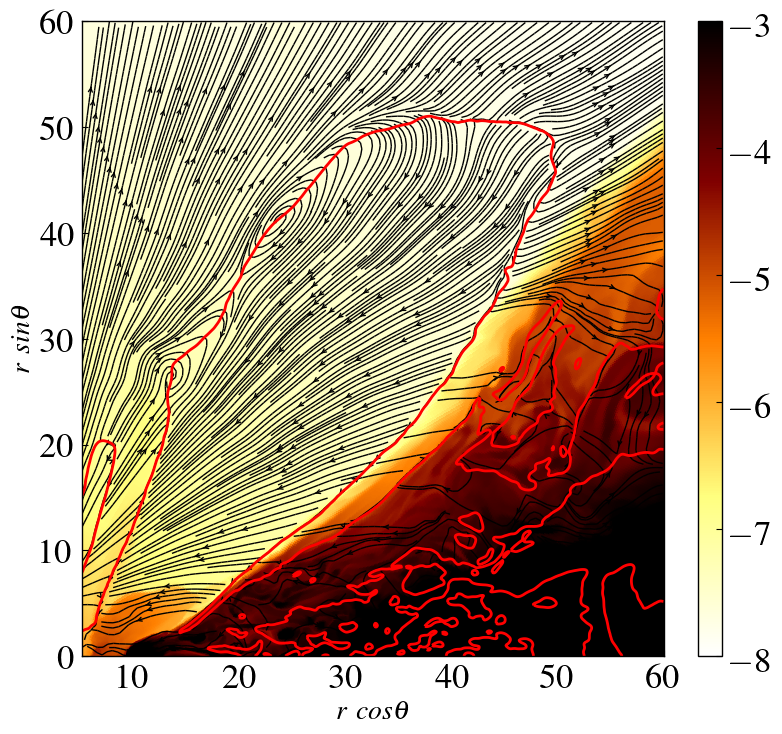}
\caption{Plot of logarithmic density profile along with the velocity streamlines for the reference model at time $t=4220$, where red line corresponds to $u^r=0$ contour.}
\label{fig-str}
\end{figure}
\begin{figure}
\centering
\includegraphics[scale=0.50]{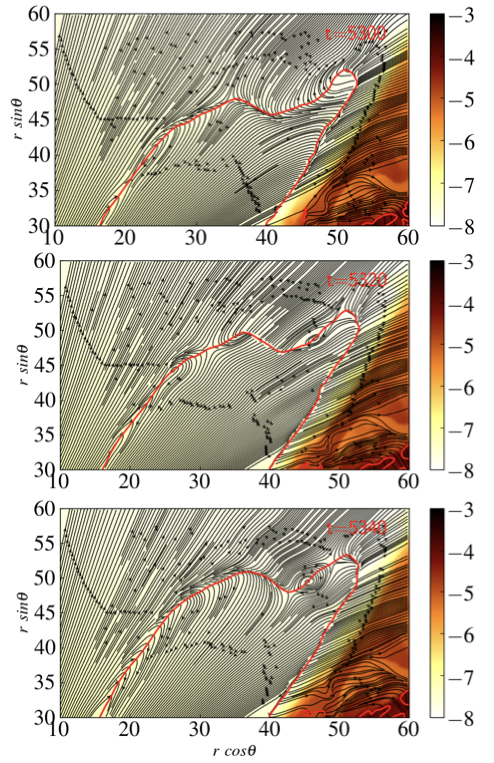}
\caption{Plot of logarithmic density profile along with the velocity streamlines for the reference model at time $t=5300, 5320,$ and $5340$ (top to bottom), where red line corresponds to $u^r=0$ contour.}
\label{fig-strtime}
\end{figure}
During the oscillation of the inner edge of the accretion disc, the inflow (high-density) and outflow (low-density) acquire shear in the transition layer. This layer may develop Kelvin–Helmholtz instability (KHI) due to shear. In Fig. \ref{fig-str}, we plot the logarithmic density profile and the velocity streamlines for the reference model at simulation time $t=4220$. The red line in the figure corresponds to the $u^r=0$ contour. The regions with $u^r<0$ and $u^r>0$ are populated with inflowing and outflowing streamlines, respectively. We observe formation of velocity vortices at the boundary layers, which is a typical signature of active KHI. The inflow-outflow region does not reach a steady-state in our simulation; instead, it continues to vary with time. In Fig.~\ref{fig-strtime}, for example, we show the temporal evolution of the density profile and velocity streamlines at three different times for the reference model ($t=5300, 5320$, and $5340$). The formation of a typical velocity vortex is depicted in the figure. The velocity vortex begins to form in the region where the flow has a high-density gradient. In addition, the inflowing and outflowing streamlines mix along the contour $u^r=0$ in the same region. These findings support the presence of KHI in the boundary layer. It should be noted that the only requirement for invoking KHI is a uniform velocity shear and no density difference (for more information, see \cite{Matsuoka2014}). As a result, the boundary layer between the relativistic jet and the inflowing matter (along the $u^r=0$ contour) is vulnerable to the formation of KHI. As a result, KHI vortices keep on forming throughout our simulation runs. The infalling matter interacts with the fast jet ejected from close to BH through KHI vortices. These vortices formed at the interface layers can facilitate mixing and also help in mass loading to the jet.
\subsection{Evolution with resistivity}
\begin{figure*}
\centering
\includegraphics[scale=0.6]{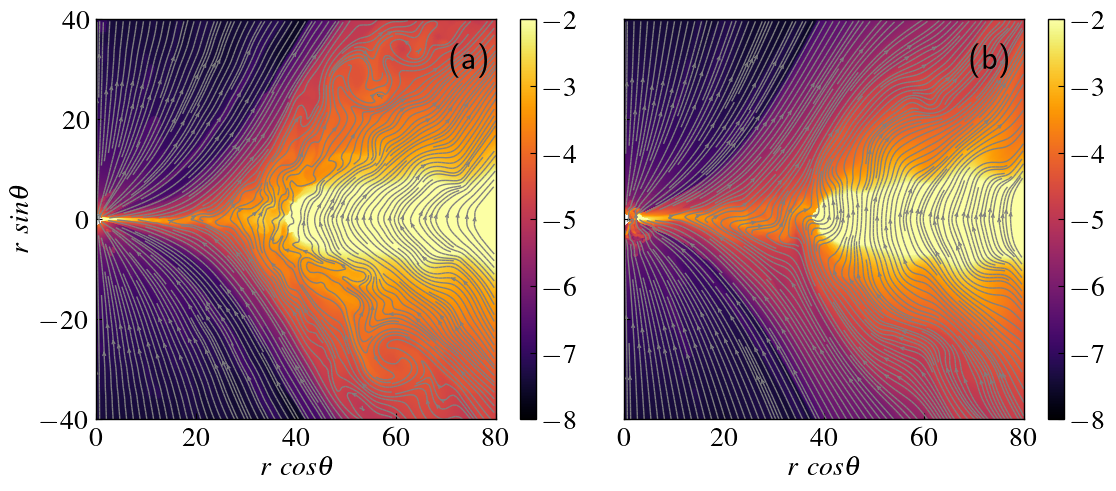}
\includegraphics[scale=0.52]{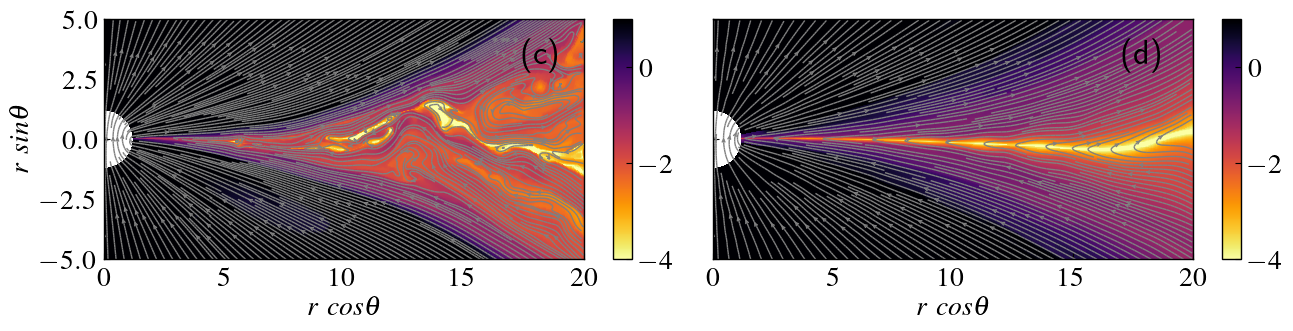}
\caption{Upper panels: Logarithmic density profile and the field lines are shown on the poloidal plane for resistive models (a) rl-2D40AH at $t=4680$ (b) rh-2D40AH at time $t=5190$.
Lower panels: Logarithmic magnetisation $(\sigma=b^2/\rho)$ profile and the field lines close to the black hole for model (c) 2D40AH (ideal MHD) at time $t=3240$ and (d) rl-2D40AH (resistive MHD) at time $t=4680$.}
\label{fig-resis}
\end{figure*}
The large-scale qualitative features of the resistive models are quite similar to those of the reference model (ideal MHD cases). We observe the accretion of matter and oscillations of the inner edge of the accretion disc within the event horizon to $r_{\rm max}$. Nonetheless, the time period of the oscillations depends on resistivity. Magnetic resistivity essentially helps in diffusing matter through the magnetic field and also impacts reconnection events in the magnetic field lines (e.g., \cite{Vourellis-etal2019,Ripperda-etal2019, Ripperda-etal2020}). In Fig. \ref{fig-resis}, we show the density profile and the magnetic field lines for our resistive models (a) rl-2D40AH ($\eta=1\times10^{-3}$) at $t=4680$ (b) rh-2D40AH ($\eta=5\times10^{-3}$) at time $t=5190$. The times are chosen such that the accreting matter is connected to the event horizon for both models. By comparing with the reference model (Fig. \ref{fig-zoom}a), for resistive cases, we do not observe the formation of plasmoids due to tearing mode instabilities. Such plasmoids form only in the domain with Lundquist number $S\gtrsim 10^{4}$, where reconnection rate is independent of $S$ and commonly known as the `fast' reconnection mode or ideal tearing mode \citep[etc.]{Bhattacharjee-etal2009, Huang-Bhattacharjee2010, Loureiro-Uzdensky2016, Striani-etal2016,  Ripperda-etal2020}. With the increase of magnetic resistivity, plasmoid instability is suppressed. A similar observation has been reported earlier by \cite{Ripperda-etal2020}. 
Also, the increase in resistivity suppresses the MRI-induced turbulence \citep[for example, see][]{Qian-etal2017}. As a result, the turbulent feature in the flow decreases with the increase of resistivity, which is evident from Fig. \ref{fig-zoom}a, \ref{fig-resis}a, and \ref{fig-resis}b.

A detailed comparison of ideal and resistive MHD flow close to the black hole is shown in Fig. \ref{fig-resis}c (ideal MHD) and \ref{fig-resis}d (resistive MHD) in terms of magnetization $(\sigma=b^2/\rho)$ and magnetic field lines. The magnetization profile is quite similar for both models, with highly magnetized $(\sigma>1)$ regions in the off equatorial plane and the matter in the equatorial plane is low magnetized $(\sigma<1)$. The figures clearly show the turbulent nature of the flow and the formation of plasmoids due to the reconnection of opposite polar magnetic field lines for the ideal model. However, in the resistive MHD model, the flow is not turbulent, and we do not observe the formation of plasmoids. Note that we consider the resistivity $\eta$ to be constant throughout the simulation domain. In general, resistivity can be a function of space and time. Recently, with such resistivity profiles, it has been shown that resistivity plays an important role in the launching process of outflows from the accretion disc in resistive GRMHD \citep{Qian-etal2018}, the generation of a turbulent outflow due to reconnection \citep{Vourellis-etal2019}, or the magnetic field amplification by a mean-field $\alpha^2$-$\Omega$ disc dynamo \citep{Vourellis-Fendt2021}.

\subsection{A transient disc structure}
\begin{figure*}
\centering
\includegraphics[scale=0.45]{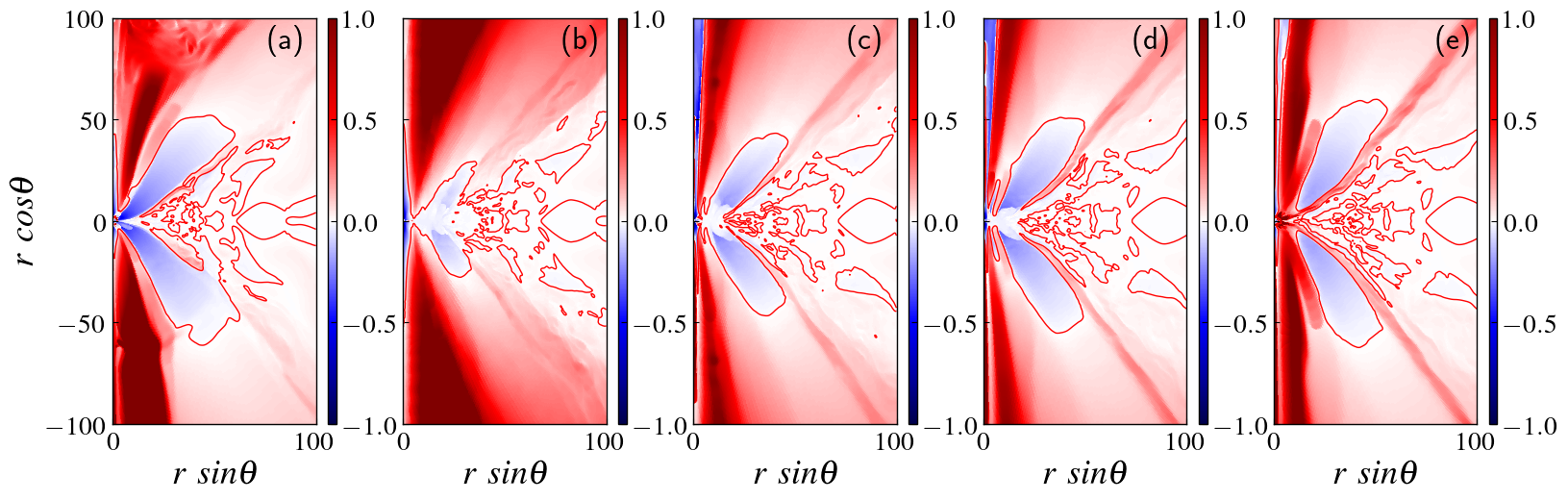}
\caption{Distribution of radial four velocity ($u^r$) for model 2D40AH at time (a) $t=3240$, (b) $t=3800$, (c) $t=4080$, (d) $t=4220$, and (e) $t=4380$. The solid red contour corresponds to $u^r=0$}.
\label{fig-ucon1}
\end{figure*}
\begin{figure*}
\centering
\includegraphics[scale=0.45]{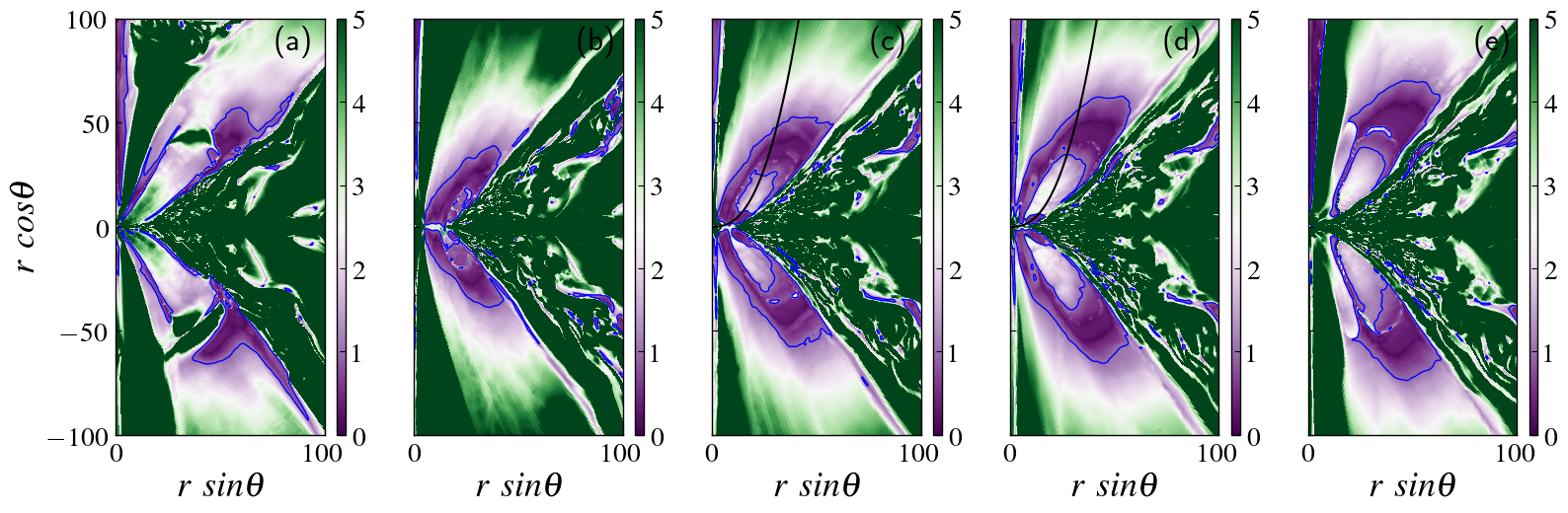}\caption{Distribution of slow-magnetosonic Mach number ($M_{\rm s}=u_p/a_s$) for model 2D40AH at time (a) $t=3240$, (b) $t=3800$, (c) $t=4080$, (d) $t=4220$, and (e) $t=4380$. Blue contour corresponds to $M_{\rm s}=1$ and the solid black line in panels c and d corresponds a parabolic streamline $y=cx^{2.5}$, with $c=0.009$.}
\label{fig-mach}
\end{figure*}
In this section, we study the slow-magnetosonic (hereafter, magnetosonic) behaviour of the truncated accretion disc. To do that, in Fig. \ref{fig-ucon1}, we show the radial four velocity ($u^r$), by following same temporal snapshots as Fig. \ref{fig-rho}.
The boundary between inflow and outflow is marked by the red contour ($u^r=0$) in Fig. \ref{fig-ucon1}. As the disc is truncated, the region with $u^r<0$ reduces and becomes fainter (see Fig. \ref{fig-ucon1}b) indicating negligible inflow, whereas the region with $u^r>0$ increases and become darker, showing a large amount of outflow. With time the region with $u^r<0$ increases, also observe an increase in the radial velocity (becomes dark blue). We also observe a sharp transition of radial velocity close to the black hole (see Fig. \ref{fig-ucon1}c-d). Such a change in velocity indicates the presence of a shock transition in the flow. Previously, many semi-analytic hydrodynamic studies have hinted at the formation of such shocks and suggested that shock solutions are viable in understanding the radiative and timing properties of astrophysical sources (\cite{Chakrabarti1989, Chakrabarti2011, Aktar-etal2015, Kumar-Indranil2017, Dihingia-etal2018, Dihingia-etal2020, Dihingia-etal2019a, Dihingia-etal2019b, Das-etal2021}, etc.). Similarly, semi-analytic MHD studies also suggested the formation of magnetosonic shock in the vicinity of the black hole \citep[e.g.,][]{Takahashi-etal2002,Takahashi-etal2006,Fukumura-etal2007}. 

\begin{figure}
\centering
\includegraphics[scale=0.6]{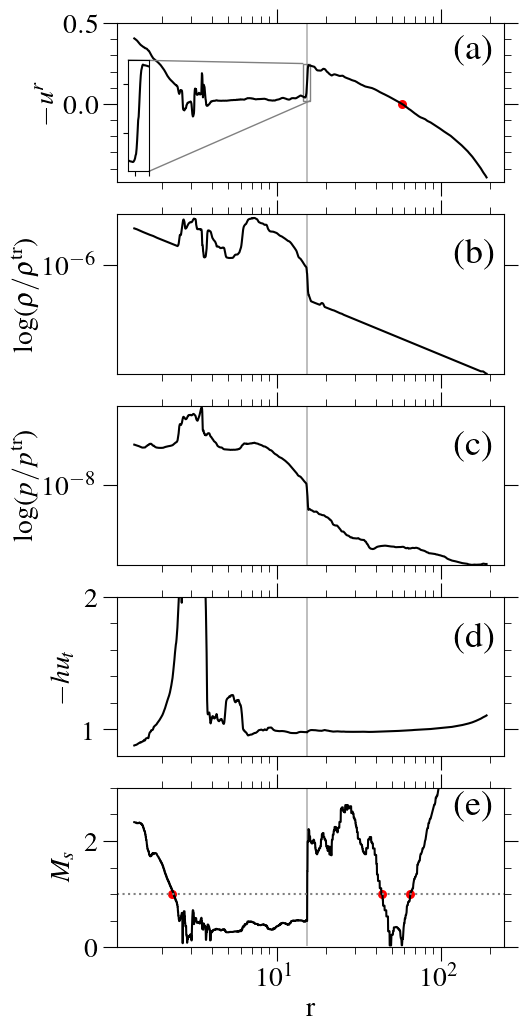}
\caption{Profile of (a) radial four velocity $(u^r)$, (b) logarithmic density $(\log(\rho/\rho^{\rm tr}))$, (c) gas pressure $(p/p^{\rm tr})$, (d) Bernoulli parameter $(-hu_t)$, and (e) Mach number ($M_s$) as a function of radial distance for model 2D40AH at time $t=4220$. The gray vertical line indicates the location of the shock transition. We show the shock transition region as an inset in panel (a). The gray horizontal dotted line in panel (e) corresponds to $M_s=1$.
The plots are shown along a parabolic streamline $y=c x^{2.5}$. See the text for more detail.}
\label{fig-shock}
\end{figure}

We show the distribution of the slow magnetosonic Mach number ($M_s=u_p/a_s$; hereafter, Mach number) by following the same temporal snapshots as Fig. \ref{fig-rho}, where poloidal velocity $u_p^2= u^ru_r + u^\theta u_\theta$, slow magnetosonic speed $a_s^2 = \frac{1}{2}(a_0^2 + a_{f}^2) - \frac{1}{2}\sqrt{(a_0^2 + a_{f}^2)^2 - 4 a_0^2 a_{f}^2 \cos^2\xi}$.
The sound speed and the poloidal Alfv\'en speed are given by $a_0^2=\gamma p/\rho h$ , $a_{f}^2= B_p^2/(B_p^2 + \rho h)$, respectively, where $B_p^2 = B_rB^r + B_\theta B^\theta$ and $h$ is the specific enthalpy of the flow. Finally, $\xi$ is the angle between the poloidal velocity vector and the poloidal magnetic field vector.
In the figure, the blue line corresponds to contour $M_s=1$. 
In Fig. \ref{fig-mach}a and \ref{fig-mach}e, we observe that accretion flow around the equatorial plane is super slow-magnetosonic (hereafter, super-magnetosonic) ($M_s>1$), where the accretion flow is connected to the event horizon. Far from the equatorial plane, we observe a region with sub slow-magnetosonic (hereafter, sub-magnetosonic) ($M_s<1$) to super-magnetosonic ($M_s>1$) flow surrounding the rotation axis of the black hole (within $\theta\sim 25^\circ - 45^\circ$ for the upper quadrant). 
In this region, very far from the black hole flow is super-magnetosonic ($M_s>1$), and also close to the black hole flow is super-magnetosonic ($M_s>1$). In between, we observe a sub-magnetosonic ($M_s<1$) region (dark violet), with a smooth transition between the layers. With time, we observe the formation of a sub-magnetosonic flow close to the black hole (Fig. \ref{fig-mach}b, see the dark violet region). As the flow moves far from the equatorial plane, the flow smoothly makes a transition from sub-magnetosonic ($M_s<1$) to super-magnetosonic ($M_s>1$) velocity. 
In Fig. \ref{fig-mach}c and \ref{fig-mach}d, we observe a super-magnetosonic region $(M_s>1)$ close to the black hole, which has a sharp transition to a sub-magnetosonic value close to the black hole $(r\sim15)$, indicating the presence of shocks in our simulations. 
For the sake of clarity, we explain this in terms of a parabolic streamline $y=cx^{2.5}$, with $c=0.009$ (depicted by solid black line in Fig. \ref{fig-mach}c and Fig. \ref{fig-mach}d). Along the streamline, very close to the black hole flow is super-magnetosonic ($r\lesssim2$), within $2\lesssim r \lesssim 15$ flow is sub-magnetosonic, and then flow is super-magnetosonic again with a shock transition (for detail see the Fig. \ref{fig-shock}e). After that, flow undergoes multiple transitions between these regimes and finally ends up as super-magnetosonic for $r\gtrsim60$.  

In Fig \ref{fig-shock}, we show the flow properties along the parabolic streamline ($y=c x^{2.5}$) for model 2D40AH at time $t=4220$. Panels (a), (b), (c), (d), and (e) correspond to the radial four-velocity $(u^r)$, density $(\rho/\rho^{\rm tr})$, pressure $(p/p^{\rm tr})$, Bernoulli parameter $(-hu_t)$ profile, and Mach number ($M_s$), respectively. The panels suggest that the flow has a shock transition in between $r_s=15.0-15.5$, shown in the figure by the gray vertical line. In Fig. \ref{fig-shock}a, we observe that at the shock, radial velocity sharply dropped from $u^r_-=0.2393$ to $u^r_+=0.0382$. Here, `-' and `+' correspond to quantities obtained before and after the shock transition. As an inset, in Fig. \ref{fig-shock}a, we show the transition of radial velocity $(u^r)$ during the shock transition.

Away from this shock transition, the radial velocity of the flow increases closer to the black hole (smaller $r$). Whereas at a larger radius far from the shock, the radial velocity becomes zero $(r \backsimeq 58.0$, see the red dot in Fig. \ref{fig-shock}a) and increases in the opposite direction as flow moves far from the black hole. This point divides the inflow from the outflow. The red contours show a similar division between inflow-outflow in Fig. \ref{fig-ucon1}. Similarly, in Fig. \ref{fig-shock}b, we observe that the density profile also shows a jump across the shock front from $\rho_-=2.13\times10^{-7}\rho^{\rm tr}$ to $\rho_+=9.08\times10^{-7}\rho^{\rm tr}$. It indicates the presence of a highly compressed shock with a density jump $R_\rho=\rho_+/\rho_-=4.26$. Similar to the density, the pressure also increases across the shock with a pressure jump $R_p=p_+/p_-\sim 10.06$ (see Fig. \ref{fig-shock}c). Consequently, the temperature jump across the shock is $R_\Theta(=R_p/R_\rho)=2.36$.
However, the specific energy or the Bernoulli parameter $(-hu_t)$ does not change significantly across the shock. This fact essentially suggests the non-dissipative nature of the shock (see Fig. \ref{fig-shock}d). Across the shock, the value of the Bernoulli parameter is less than unity $(-hu_t\sim0.98)$, indicating that the immediate post-shock flow is gravitationally bound $(-hu_t<1)$.

As evident from panel (d) of Fig. \ref{fig-mach}, in Fig. \ref{fig-shock}e, the Mach number ($M_s$) profile along the streamline shows three magneto-sonic points at $r= 2.3, 43.9,$ and $65.0$, respectively.
These critical points are denoted as red dots in Fig. \ref{fig-shock}e. The gray horizontal line in Fig. \ref{fig-shock}e corresponds to $M_s=1$, indicating the division between the sub-magnetosonic and super-magnetosonic regions. For the outgoing branch $(u^r>0, r>58.0)$, the sub-magnetosonic flow becomes super-magnetosonic after crossing the magnetosonic points at $r=65.0$. The in-going branch $(u^r<0, r<58.0)$ becomes super-magnetosonic after crossing the magnetosonic point at $r=43.9$ but becomes sub-magnetosonic after the shock transition $(r_s)$. Finally, the flow becomes super-magnetosonic after crossing the magnetosonic point at $r=2.3$. We observe that Mach number jumps from $M_{s,-}=1.98$ to $M_{s,+}=0.49$ across the shock front.

\begin{figure}
\centering
\includegraphics[scale=0.42]{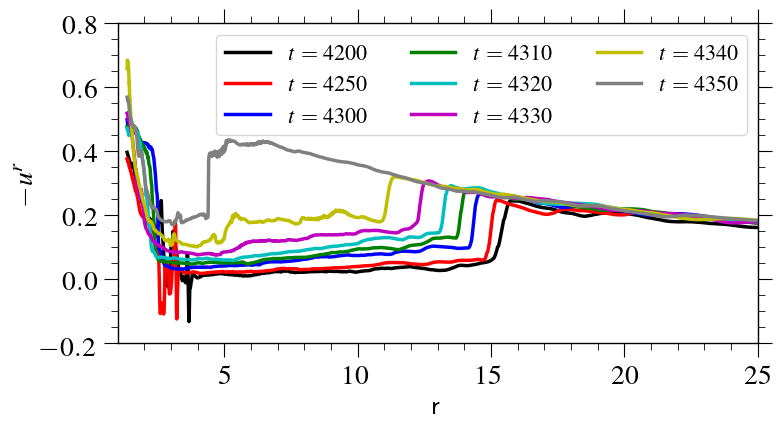}
\caption{Profile of radial velocity for 2D40AH from time $t=4200$ to $t=4350$, time corresponding to different curves are labeled on the figure. The plots are shown along a parabolic streamline $y=c x^{2.5}$. See the text for more detail.}
\label{fig-shockdy}
\end{figure}

To understand the dynamics of the shock, in Fig. \ref{fig-shockdy}, we show the temporal evolution of the radial velocity profile for 2D40AH from time $t=4200$ to $t=4350$. The shock locations obtained at times $t=4200, 4250, 4300, 4310, 4320, 4330, 4340$, and $t=4350$ are $r_s=15.40, 14.98, 14.32, 13.83, 13.20, 12.33, 11.09$ and $r_s=4.44$, respectively (here, we consider the mean value). Thus, shock is not steady and the shock front moves towards the black hole with time. The velocity of the shock front also increases with time. Finally, the shock disappears with a huge outflow. After the outflow, as advection starts to dominate, shock appears again. We observe the appearance and disappearance of shock throughout the simulation time in all the simulation models. This essentially suggests that the shock transition in the truncated accretion disc is a transient phenomenon. 

In summary, we observe that the flow creates a hot, comparatively high-density region around the black hole during the shock transition. Also, note that the density of this post-shock region is much lower than that of the thin disc. This hot region can be associated with an extended corona surrounding the black hole. Many earlier authors anticipated the formation of such an extended corona from the truncated disc \citep{Eardley-etal1975, Ichimaru1977}. With temporal evolution, the size of the corona also changes as the shock front moves towards the black hole. The variations in the size of the corona may lead to fluctuations in the electromagnetic emission. As a result, these fluctuations in emission could be useful in understanding variability in astrophysical sources.

\begin{figure*}
\centering
\includegraphics[scale=0.4]{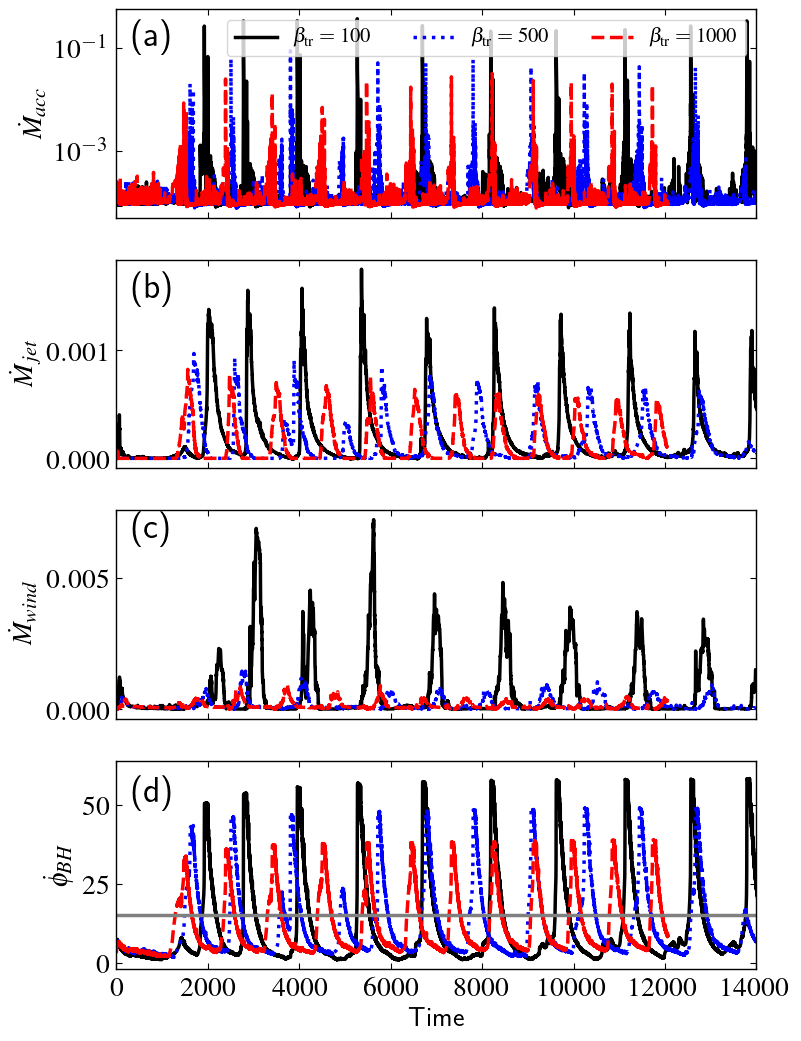}
\includegraphics[scale=0.4]{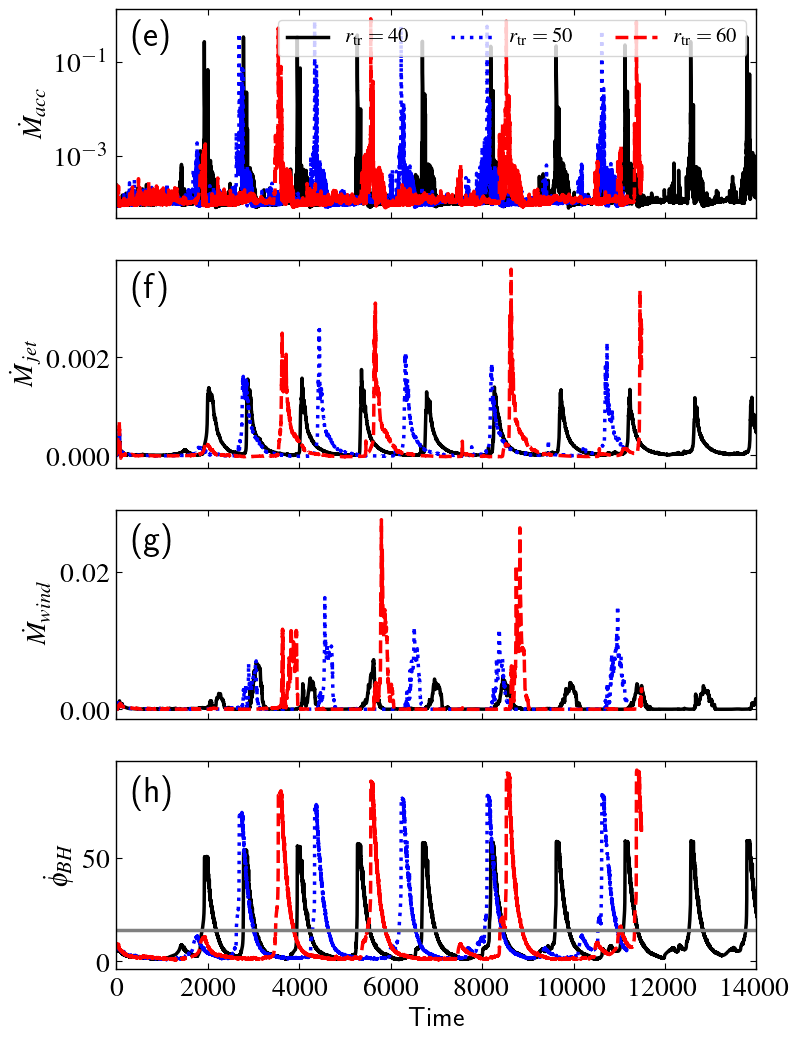}
\caption{Profiles of accretion rate $(\dot{M}_{\rm acc})$, mass flux through funnel $(\dot{M}_{\rm jet})$, mass flux through wind  $(\dot{M}_{\rm wind})$, and magnetic flux accumulated at the event horizon  $(\dot{\phi}_{\rm BH})$ for models with different $\beta_{\rm tr}$ and $r_{\rm tr}$. In panels a-d (left), plots are shown for models 2D40A, 2D40B, and 2D40C, corresponding to $\beta_{\rm tr}=100, 500,$ and $1000$, respectively, with $r_{\rm tr}=40$. In panels e-h (right), plots are shown for models 2D40A, 2D50, and 2D60, corresponding to $r_{\rm tr}=40, 50,$ and $60$, respectively, with $\beta_{\rm tr}=100$. The gray horizontal line in panels (d) and (h) correspond to $\dot{\phi}_{\rm BH}=15$.}
\label{fig-fluxes}
\end{figure*}

\section{Temporal characteristics}
In this section, we study the long-term temporal evolution of the truncated accretion disc in terms of integrated quantities and compare their properties for different simulation models (ideal MHD).
The mass accretion rate $(\dot{M}_{\rm acc})$, mass flux through funnel $(\dot{M}_{\rm jet})$, mass flux through wind  $(\dot{M}_{\rm wind})$, and magnetic flux accumulated at the event horizon  $(\dot{\phi}_{\rm BH})$ are shown in Fig. \ref{fig-fluxes} for different simulation models. The $\dot{M}_{\rm acc}$ is calculated at the event horizon, and $\dot{M}_{\rm jet}$ and $\dot{M}_{\rm wind}$ are calculated at a radius $r=50$ following \cite{Porth-etal2017, Nathanail-etal2020}.
Note that, in $\dot{M}_{\rm jet}$, we only consider highly magnetised, relativistic outflow with $\sigma>1, -hu_t>1$, or efficiency factor of the Poynting flux $\xi>2$, where $\xi=(-T^r_t - \rho u^r)/\rho u^r$. On the other hand, in $\dot{M}_{\rm wind}$, we consider outflow with $\sigma<1, -hu_t>1$, or efficiency factor of the Poynting flux $\xi<2$.
In the left panels, we study the role of initial plasma-$\beta$ parameters ($\beta_{\rm tr}$), and in the right panels, we investigate the role of truncation radius on these quantities.

\begin{figure}
\centering
\includegraphics[scale=0.45]{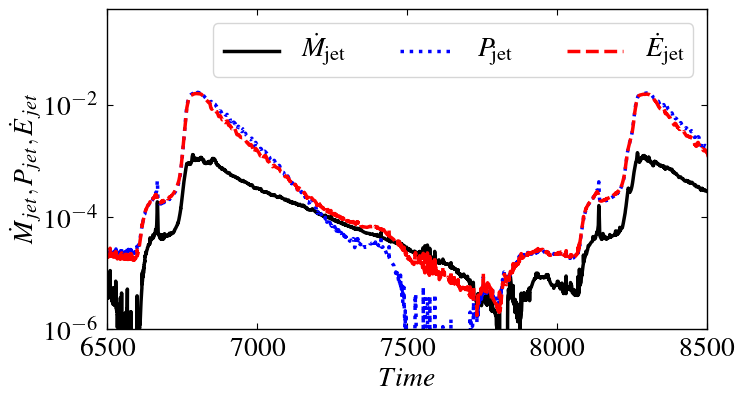}
\caption{Profile of $\dot{M}_{\rm jet}$, total energy flux ($P_{\rm jet}$) and electromagnetic energy flux ($\dot{E}_{\rm jet}$) through the jet in code unit as a function of simulation time for model 2D40A.}
\label{fig-jet}
\end{figure}

With the transport of angular momentum, the matter in the inner edge sets the accretion process. Once the accreted matter reaches the event horizon, we observe a spike in the accretion rate $(\dot{M}_{\rm acc})$ profile (see Fig. \ref{fig-fluxes}a). The matter drags the magnetic field lines along with it, and we observe magnetic field lines rooted to the ergo-sphere. Such a magnetic field structure is required to have an active Blandford-Znajek (BZ) process \citep{Blandford-Znajek1977, Komissarov-Barkov2009}. To better understand the properties of relativistic jet, in Fig. \ref{fig-jet}, we show the total energy ($P_{\rm jet}$) flux and electromagnetic energy flux ($\dot{E}_{\rm jet}$) through the jet along with the $\dot{M}_{\rm jet}$ in code unit for model 2D40A from $t=6500$ to $t=8500$. $P_{\rm jet}$ and $\dot{E}_{\rm jet}$ are calculated following \citep{McKinney-etal2012,Nathanail-etal2020},
\begin{align}
    P_{\rm jet}= \int \left(-T^r_t -\rho u^r\right)\sqrt{-g} d\theta d\phi,
\end{align}
\begin{align}
    \dot{E}_{\rm jet} = \int -{T^{EM}}^r_t\sqrt{-g} d\theta d\phi, 
\end{align}
where, superscript EM denotes only the electromagnetic part of the energy-momentum tensor (for detail see \cite{McKinney-etal2012}. The integration is performed only in the funnel region at $r=50$, with $\sigma>1, -hu_t>1$ or $\xi>2$. We observe peak value of $\dot{M}_{\rm jet}$, $P_{\rm jet}$, and $\dot{E}_{\rm jet}$ at the same time (see Fig. \ref{fig-jet}), suggesting the high-mass flux through the funnel is Poynting dominated and highly relativistic. These facts strongly suggest that the active BZ process results in a higher value of $\dot{M}_{\rm jet}$ (see Fig. \ref{fig-fluxes}b).
Note that the density in the funnel region is very low and depends on the density floor model (see the black color in Fig. \ref{fig-rho}). 

\begin{figure}
\centering
\includegraphics[scale=0.45]{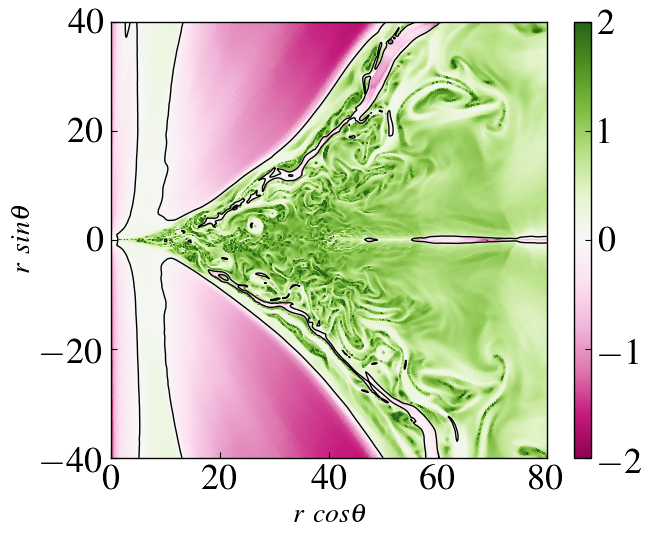}
\caption{Distribution of logarithmic ratio between the toroidal $(\sqrt{B_\phi B^\phi})$ component to the poloidal component $(B_p)$  of the magnetic field for the reference model at a simulation time $t=3240$. The black solid line illustrates the contour  $\sqrt{B_\phi B^\phi}/B_p=1$.}
\label{fig-bphibz}
\end{figure}

Similarly, to understand the peaks in the disc-wind profiles (Fig. \ref{fig-fluxes}c), in Fig. \ref{fig-bphibz}, we show the ratio of the toroidal $(\sqrt{B_\phi B^\phi})$ component to the poloidal component $(B_p=\sqrt{B^rB_r + B_\theta B^\theta})$ of the magnetic field for the reference model at a simulation time $t=3240$. The black solid line in the figure represents the boundary of the $\sqrt{B_\phi B^\phi}/B_p=1$ contour. Depending on the contribution of these two components the disc wind could be Blandford-Payne type wind or $(B_{\phi})$ dominated disc-wind \citep[for discussion see][]{Dihingia-etal2021}. The figure essentially suggest that the high-density matter in the equatorial plane has very strong toroidal component of magnetic field as compared to the poloidal component of the magnetic field $(\sqrt{B_\phi B^\phi}/B_p>1)$. The values of the ratio on the equatorial plane ranges $\sqrt{B_\phi B^\phi}/B_p\sim 10-100$. Thus, the matter close to the black hole is subjected to a gradient of pressure due to the toroidal component of the magnetic field, resulting in a strong toroidal magnetic field $(B_{\phi})$ dominated disc-wind.
Consequently, we observe a higher value of $\dot{M}_{\rm wind}$ in this stage (see Fig. \ref{fig-fluxes}c).

In the highly accreting stage, magnetic flux accumulates around the event horizon much faster. Once the magnetic flux accumulated around the horizon reaches $\dot{\phi}_{\rm BH} > 15$ (see the gray horizontal line), flow achieves a magnetically arrested disc (MAD) (\cite{Tchekhovskoy-etal2011}, see Fig. \ref{fig-fluxes}d). As discussed in section 3.1, magnetic flux does not keep on building around the black hole, and eventually it loses the magnetic flux via an eruption event. 
The eruption events give rise to strong flares in jet rate as well as in wind rate (see Fig. \ref{fig-fluxes}b-c). Such eruption events are very helpful in understanding the observed near-infrared (NIR) flares from Sgr A* \citep{ Dexter-etal2020, Chatterjee-etal2020a, Porth-etal2021}. After the eruption event, the accretion disc becomes truncated again and the accretion rate at the event horizon  ($\dot{M}_{\rm acc}$)
drops to its lowest value ($\dot{M}_{\rm acc}\lesssim10^{-4}$, see Fig. \ref{fig-fluxes}a). Due to the lack of matter close to the black hole, the mass flux through the jet and wind also becomes negligible (see Fig. \ref{fig-fluxes}c-d). 
Subsequently, the magnetic flux accumulated around the horizon becomes $\dot{\phi}_{\rm BH}< 15$. Over time, flow again accumulates magnetic flux around the horizon and becomes $\dot{\phi}_{\rm BH}> 15$ (see Fig. \ref{fig-fluxes}d). We observe the quasi-periodic oscillation of the magnetic flux throughout our simulations. Consequently, the other mass flux rates $(\dot{M}_{\rm acc}, \dot{M}_{\rm jet},~{\rm and}~ \dot{M}_{\rm wind})$ also show peak values in a similar manner.
We observe a time lag in the peak value in accretion rate $(\dot{M}_{\rm acc})$, mass flux through funnel $(\dot{M}_{\rm jet})$, and mass flux through wind  $(\dot{M}_{\rm wind})$ profiles. The time taken by the jet and the disc-wind from the launching site to reach radius $r = 50$ appears as the observed time lag in Fig. \ref{fig-fluxes} (for details see Fig. 15 of \cite{Dihingia-etal2021}).

\begin{figure*}
\centering
\includegraphics[scale=0.5]{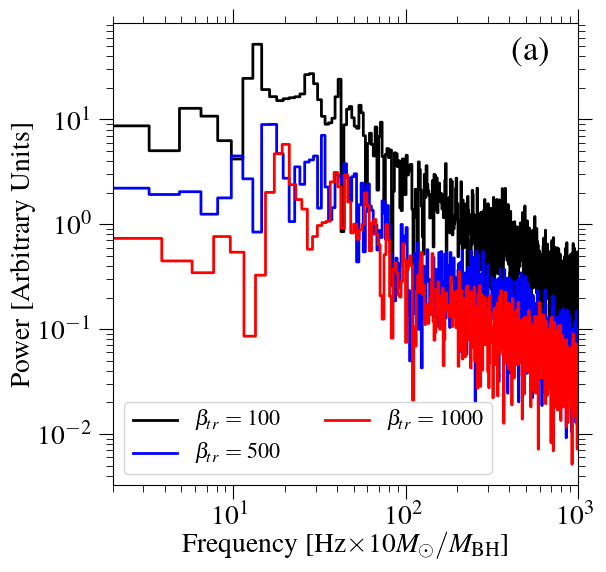}
\includegraphics[scale=0.5]{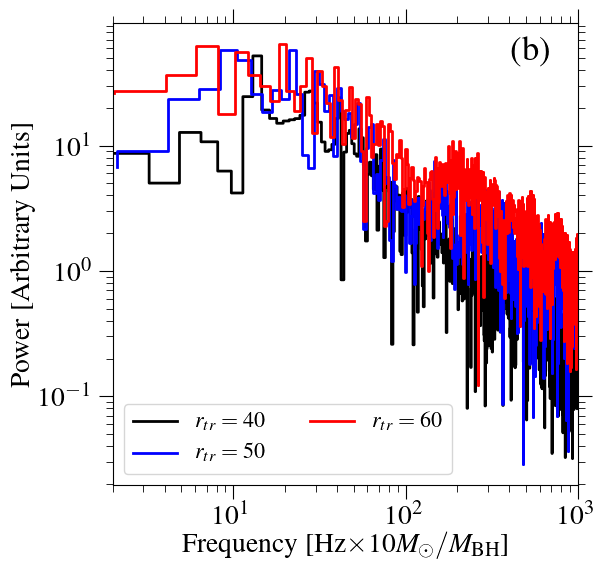}
\caption{PDS corresponding to the accretion rate profile for different {simulation models with different values of (a) initial plasma-$\beta$ ($\beta_{\rm tr}$) parameters and (b) truncation radius $(r_{\rm tr})$. In the panel (a) black, blue, and red lines correspond to $\beta_{\rm tr}=100, 500$, and $1000$, respectively. In the panel (b) black, blue, and red lines correspond to $r_{\rm tr}=40, 50$, and $60$, respectively.}}
\label{fig-PDS}
\end{figure*}

To understand the statistical behaviour of the these oscillations, we plot the power density spectrum (PDS) of the accretion rate profile $(\dot{M}_{\rm acc}$) for different simulation models at Fig. \ref{fig-PDS}. In panel Fig. \ref{fig-PDS}a, variation of PDS is shown for different values of $\beta_{\rm tr}$ and in panel Fig. \ref{fig-PDS}b, the same is shown for different values of truncation radius. In the figure, the power is plotted in an arbitrary unit while the frequency is plotted in Hz by converting the simulation time to the physical unit (seconds). Although the peaks in the accretion rate profile are not statistically rich due to the limited run time of the simulation $t\sim12000-14000$ code unit or $t\sim0.6-0.7(M_{\rm BH}/10M_{\odot})$ second in physical unit. We observe a frequency corresponding to maximum power in the PDS $(\nu_{\rm QPO, max})$. This fact essentially suggests the quasi-periodic nature of the oscillations.

The strength of the magnetic field is the primary driver of activities in the accretion flow. With the increase of magnetic field strength (lower $\beta_{\rm tr}$), the magnetic tension force due to the reconnected field lines also increases. Stronger ram pressure is required to penetrate the stronger magnetic tension force. Consequently, the accreting matter needs more time to reach the event horizon, and as a result, the time period of oscillation increases with the strength of the magnetic field (lowering of  $\beta_{\rm tr}$). For the models with $\beta_{\rm tr}=100$, $500$, and $1000$, the $\nu_{\rm QPO, max}$ of the PDS are $\nu_{\rm QPO, max}\sim 13.5$, $16.0$, and $20$ $Hz\times (10 M_{\odot}/M_{\rm BH})$, respectively.

With the increase of the truncation radius $(r_{\rm tr})$, the accretion starts from a larger distance from the black hole and with a higher value of specific angular momentum. The flow loses the excess angular momentum via different channels (e.g., MRI, disc-winds) and moves towards the black hole. 
Therefore, the infall time of the matter at the inner edge of the truncation radius increases with the truncation radius. 
Consequently, mass accretion between the event horizon and truncation radius takes longer for models with a higher truncation radius. As a result, the accumulation of magnetic flux happens at a much lower rate. Eventually, the time period of oscillation increases with truncation radius. For the models with $r_{\rm tr}=40$, $50$, and $60$, the $\nu_{\rm QPO, max}$ of the PDS are $\nu_{\rm QPO, max}\sim 13.5$, $9.5$, and $7.0$ $Hz\times (10 M_{\odot}/M_{\rm BH})$, respectively. 

\begin{figure}
\centering
\includegraphics[scale=0.4]{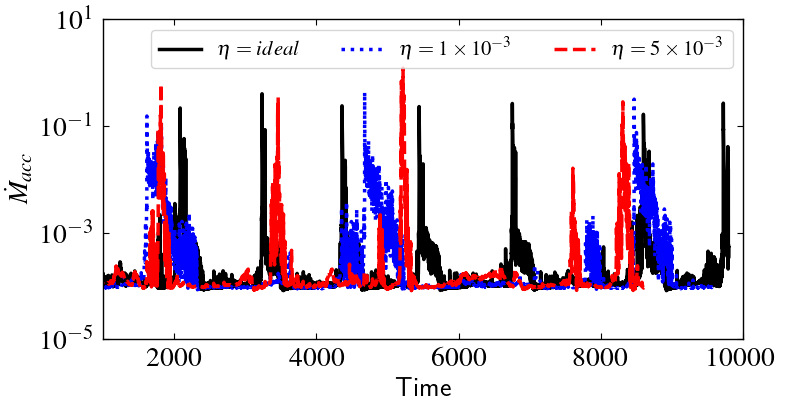}
\caption{Plot of temporal evolution of accretion rate profile for models with different resistivity, $\eta=0$ (black), $1\times 10^{-3}$ (blue), and $5\times 10^{-3}$ (red). }
\label{fig-comp}
\end{figure}
To study the role of resistivity in the dynamical properties of the truncation disc, we plot accretion rate as a function of simulations time for models 2D40AH (ideal MHD), rl-2D40AH ($\eta=1\times10^{-3}$), and rh-2D40AH ($\eta=5\times10^{-3}$) in Fig. \ref{fig-comp}a. The figure shows that the temporal evolution of the accretion rate is different for models with different resistivity. In resistive flow, the matter can diffuse through the magnetic field without preserving the frozen-in condition as in the ideal MHD. In such a situation, comparatively lower ram pressure is required to penetrate the magnetic tension force in the disc mid-plane. As a result, with the increase in magnetic resistivity of the flow, the time period of oscillation decreases. As a result, the QPO frequency $(\nu_{\rm QPO, max})$ is expected to increase with the increase in magnetic resistivity.

However, in the ideal MHD case, due to the active `fast' reconnection mode, we observe a much shorter time period of oscillations as compared to resistive models. For the same reason, the accretion peaks are sharper than those of the other models. We also note that, at time $t\sim9000$, radius of balancing magnetic tension force and the ram pressure for magnetic resistivity $\eta=1\times10^{-3}$ is $r_{\rm max}\sim37$. Subsequently, for $\eta=5\times10^{-3}$ the radius of balancing magnetic tension force and the ram pressure decreases to $r_{\rm max}\sim33$. 

We see sharp peaks in the mass and magnetic flux rates in Figs.~\ref{fig-fluxes} and \ref{fig-comp}. 
We stress that these simulations are carried out in an axisymmetric framework. To compare these results to a three-dimensional (3D) evolution, we ran a 3D simulation using the same setup as the reference run, but with a modest resolution of $224\times96\times128$. 
Clearly, the resolution of our ideal MHD 3D run is insufficient to capture some critical details of the system, such as magnetic field line reconnection. This exemplary 3D simulation, on the other hand, helps us understand the reality of features observed in the accretion flow \citep[e.g.,][]{Porth-etal2021,Ripperda-etal2022}. 
We do not see such sharp peaks in the mass and magnetic flux profiles in 3D unlike the ideal axisymmetric runs.
Instead, we see broader peaks caused by the formation of accretion {\it fingers} (for a detailed discussion, see Appendix B).

\section{Radiative characteristics}
Radiative transfer calculations are essential to correlate our simulations with the astrophysical observations in the electromagnetic paradigm. In this section, we study the radiative signature of the truncated accretion disc during the oscillation of inner accretion flow.
\begin{figure}
\centering
\includegraphics[scale=0.45]{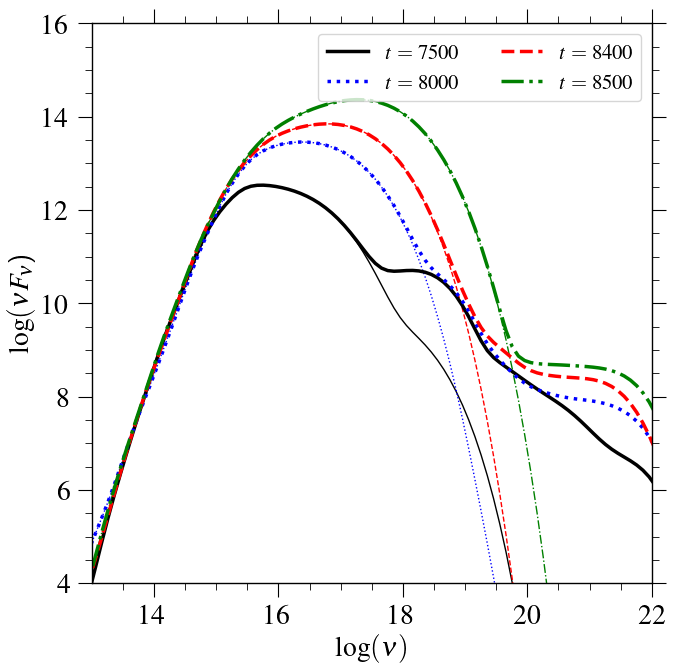}
\caption{Emission spectrum for model 2D40AH at time $t=7500, 8000, 8400$, and $t=8500$, marked on the figure.
Thinner lines correspond to the emission spectrum only due to the thermal synchrotron process.}
\label{fig-spec}
\end{figure}
\begin{figure}
\centering
\includegraphics[scale=0.45]{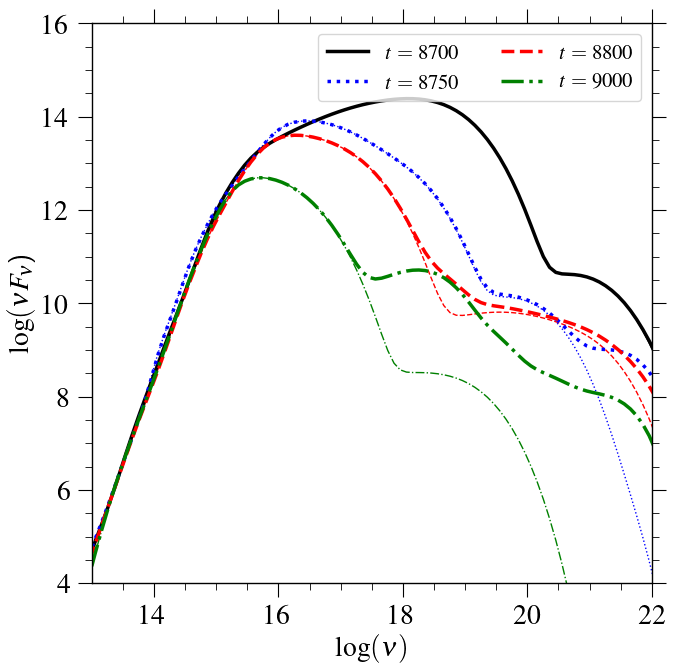}
\caption{Emission spectrum for model 2D40AH at time $t=8700, 8750, 8800$, and $t=9000$, marked on the figure. Thinner lines correspond to the emission spectrum only due to the thermal synchrotron process.}
\label{fig-specr}
\end{figure}
\subsection{Post-processing tool and scaling}
We incorporate the GRRT post-processing module \texttt{RAPTOR} and calculate the near horizon emission from the reference run. \texttt{RAPTOR} reproduces the appearance and spectrum of black hole sources to a distant observer considering the effects due to a strong gravitational field (viz. gravitational lensing, redshift, and relativistic beaming) \citep{Bronzwaer-etal2018}. \texttt{RAPTOR} also allows us to incorporate different radiative processes to render emission from the accretion disc.

In this study, we aim to study emission signatures in the hard-intermediate state. Accordingly, we consider the thermal synchrotron and Bremsstrahlung processes as the sources of emission and neglect the black body component, which may be necessary for the soft and soft-intermediate states of an outburst.
We calculated the spectrum for a black hole with a mass $M=10M_{\odot}$. To model the electron temperature from the flow temperature, we use $R-\beta$ prescription following \cite{Moscibrodzka-etal2016}, where we choose $R_l=1, R_h=60$ following \cite{Mizuno-etal2021}. We collected all the emissions on a screen from $r\lesssim 60$ and fixed the line of sight angle at $i=60^{\circ}$. We fixed the distance of the screen at $r_{\rm cam}=10^4r_g$. To mimic the accretion flow around a BH-XRB, we scale the density in cgs units with the help of the rest-mass density scaling factor $\rho_{\rm unit}=M_{\rm unit}/r_g^3$, where we consider $M_{\rm unit}=10^{13}$ gm . Accordingly, we scale energy density, magnetic field strength, and number density in cgs units using, $U_{\rm unit}=\rho_{\rm unit}c^2$, $B_{\rm unit}=c\sqrt{4\pi\rho_{\rm unit}}$, and $N_{\rm unit}= \rho_{\rm unit}/(m_e + m_p)$, respectively, where $m_e$ and $m_p$ are mass of electron and proton. 

With this, the accretion rate calculated when the inner edge is attached and detached to the event horizon is of the order of $\dot{M}_{\rm acc}\sim10^{-3}$ and $\dot{M}_{\rm acc}\sim10^{-5}$ Eddington units ($\dot{M}_{\rm Edd}=1.44\times10^{17} \left(M/M_{\odot}\right)$ g s$^{-1}$), respectively. Note that the spectral properties of an accretion disc depend on the mass of the black hole, accretion rate, line of sight angle, non-thermal particles, $R_l$, $R_h$ parameters, etc. \citep[e.g.,][]{Bandyopadhyay-eyal2021,Mizuno-etal2021,Fromm-etal2021}. For this study, we do not intend to do parametric studies. Instead, we want to study the qualitative properties of emission during the oscillation of the inner edge of the accretion disc. Therefore, we only considered one set of parameters throughout the study.

\subsection{Synthetic emission spectra}
In Fig. \ref{fig-spec}, we plot the emission spectrum for the reference run at different simulation times, where the thin and thick lines present the contribution only from the thermal synchrotron process alone and that including Bremsstrahlung, respectively. In the figure, $\nu F_\nu$ is expressed in units of `Jy~Hz'. Different simulation times are marked on the figure. During this simulation time range, the accretion matter moves from the radius $r_{\rm max}$ to the event horizon. The emission spectra show a peak value around $\nu\sim 10^{16}-10^{18}Hz$, which corresponds to the emission due to the thermal synchrotron process and dominates up to $\nu\lesssim10^{19}Hz$.  Whereas the high energy $(\nu\gtrsim 10^{19})$ part of the spectrum is dominated by the emission due to the Bremsstrahlung process. 

The emissions from the Bremsstrahlung process show a distinct shape change with respect to frequency within $\nu\sim10^{18-20}$Hz, which signifies the presence of regions with a sharp change in temperature and density. At the time $t=7500$ (for solid black line), the accretion disc is truncated, and the inner edge of the thin disc is also far from the black hole. Overall, the flow is cold, and the emission comes only from the thin Keplerian disc and weak disc-winds. Consequently, we observe lower emissions in the high-energy region. 

As the flow gradually loses angular momentum, the inner edge of the disc approaches the black hole. During this process, the matter in the inner edge becomes hot, accumulates more magnetic flux, and disc-winds also become stronger. This results in higher intensity of thermal synchrotron and Bremsstrahlung emission. Also, we observe that peak due to thermal synchrotron emission moves towards the higher frequency range as matter approaches the black hole. As the flow reaches close to the event horizon $(t=8500$, see the green dot-dashed line), the emission within energy range $(h\nu \gtrsim 0.1KeV, \sim 10^{16}{\rm Hz})$ increases by two orders of magnitude. In comparison, the lower energy part remains unaltered, as the low magnetized outer part of the disc remains steady during this period. However, the emission due to the Bremsstrahlung increases only by one order of magnitude during this process.

To understand the radiative properties during the eruption events, in Fig. \ref{fig-specr}, we plot the emission spectrum for different simulation times (marked on the figure) when the inner edge of the disc is receding from the black hole. The thin and thick lines in the figures are drawn following a similar convention as in Fig. \ref{fig-spec}. In the figure, we observe the opposite trend to that of Fig. \ref{fig-spec}. As the truncated disc recedes, it shows strong outflow in terms of jet and disc-wind (see Fig. \ref{fig-fluxes}). With the outflow, accretion flow loses the advected magnetic flux. Consequently, the magnetic field strength decreases, and we observe that the thermal synchrotron emission decreases with time. Also, the synchrotron peak shifts towards the lower energy range. 

We observe that most of the emission is contributed by the thermal synchrotron process during the receding, even in the high energy region $(\nu\gtrsim10^{19}$Hz), signifying a very strong magnetic field around the black hole. The Bremsstrahlung starts to dominate again in the high energy range as the disc reaches far from the black hole ($t=9000$, see the green dot-dashed line). By then, the disc-winds leave our region of interest ($r\lesssim60$) and do not contribute to the emission spectra.

\begin{figure*}
\centering
\includegraphics[scale=0.52]{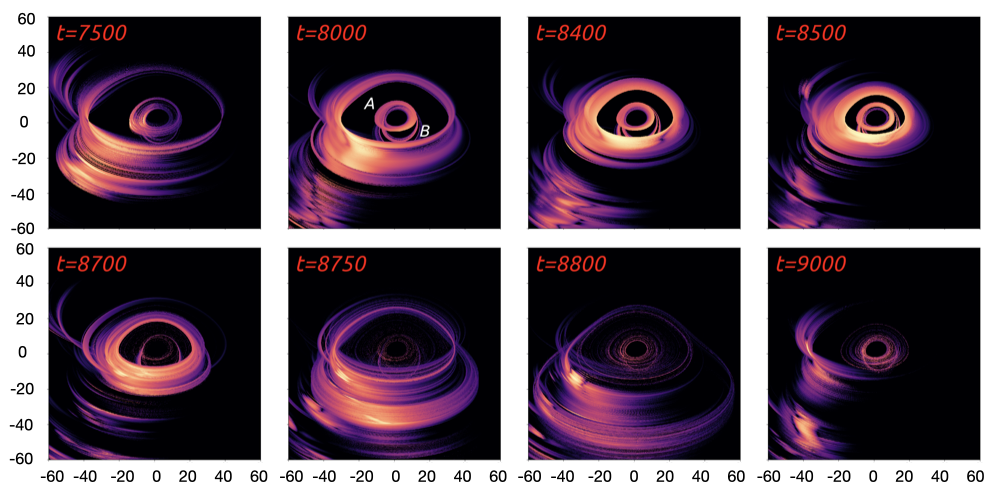}
\caption{Intensity maps at $1KeV$ energy for simulation times same as Fig. \ref{fig-spec} and \ref{fig-specr} (marked on the figure). The domain of the figures is expressed in units of $r_g$.}
\label{fig-map1k}
\end{figure*}
\subsection{Synthetic intensity maps}
The 2D synthetic intensity at $h\nu=1KeV$ for the reference run at different simulation times (same as Fig. \ref{fig-spec} and \ref{fig-specr}) is shown in Fig. \ref{fig-map1k}. In the figures, we plot the normalized intensity in logarithmic scale within a box  $x:[-60, 60]r_g$ and $y:[-60, 60]r_g$. In the figure, we broadly observe two concentric rings. The size of the outer ring is about $40r_g$ and the inner is about $10r_g$. These maps essentially suggest that the inner edge of the truncated accretion disc and the disc-wind contribute to most of the emission, which constitutes the outer ring. As the inner edge moves towards the event horizon $t\sim 7500-8500$, the radius of the brightened edge decreases. At the time, $t=8000, 8400$, and $8500$, we observe two extra ring-like structures surrounding the black hole due to the hot and dense post-shock matter. The post-shock region exists on both sides of the equatorial plane. Thus, the emission coming out of the post-shock region appears as two concentric rings around the black hole to the observer. The rings at time $t=8000$ are marked on the figure as `A' and `B'. As the post-shock regions on both sides of the equatorial plane appear at slightly different inclinations to the observer. As a result, the inner rings are located more asymmetrically with respect to the equatorial plane than the outer ring.

Unlike the standard image of the black hole \citep{EHTI-2019}, here we observe a multiple ring structure around the event horizon. As the inner edge of the disc recedes outwards ($t\sim8700-9000$), we observe negligible emission from the low-density flow close to the black hole. During this time, most of the emission is coming from the disc wind. As the wind leaves our area of interest $(r\lesssim60)$, total emission drops drastically ($t=9000$). Our observation of an extended ring structure nicely fits with the GR-MHD ray-traced emission maps by \citet{Bandyopadhyay-eyal2021}, who investigated how the structural components of a BH jet source - the BH, the disc, the spine jet, and the disc wind - may be disentangled by their radiative imprints.

Due to the transport of angular momentum, the sub-Keplerian matter accretes towards the black hole. Also, some of this accreting matter contributes to the outflow as jet and disc-wind. This indicates that the low angular momentum matter around the black hole plays a crucial role in the emission of hard X-rays. Consequently, the emissions coming within the energy range $h\nu\sim1-100keV (\nu\sim 10^{17}-10^{19}Hz)$ change in two-orders of magnitude during the oscillation (see Fig. \ref{fig-spec}). Similarly, in the high energy range  ($h\nu>100KeV$ or $\nu >10^{19}Hz)$, the emission increases by one order of magnitude in the oscillating phase. These features may significantly impact the understanding of spectral and timing properties in BH-XRBs. BH-XRBs often exhibit spectral variability in their outbursting phase (\cite{Belloni2010, Belloni-etal2011, Belloni-Motta2016, Ingram-Motta2019}, etc.).

Note that in this study, we do not consider any Comptonization process in the calculation of the emission features. However, it is needless to mention that the inclusion of the Comptonization process is essential to understand the hard X-rays from BH-XRBs (e.g., \cite{Steiner-etal2009, Titarchuk-etal2014, Poutanen-etal2018}, etc.). With the inclusion of the Comptonization process, the high-energy emissions from the hot disc-wind and the post-shock region are expected to increase. 

The radiation and spectral signatures calculated due to the thermal synchrotron and Bremsstrahlung depend on the considered parameters of the $R-\beta$ relation. In reality, a large number of these parameters $(R_l, R_h)$ need to be tested to find a suitable combination/range of parameters that match the observations \citep[e.g.,][]{Mizuno-etal2021,Fromm-etal2021}. Such scaling relations are often used in literature to calculate the near horizon radiative properties of AGNs \citep[e.g., ][]{EHTI-2019}. A more consistent approach would be to account for two-temperature fluid and include relevant heating/cooling terms to obtain the electron temperature and the emission self-consistently \citep[e.g.,][]{Ressler-etal2015, Sadowski-etal2017, Ryan-etal2017, Mizuno-etal2021,Dihingia-etal2022}. Nevertheless, with the inclusion of these processes, the qualitative radiative feature of modulation of emission during the oscillation of accretion flow is expected to be the same.

\section{Summary and discussion}
In this work, we set up a highly resolved, magnetized, truncated accretion disc in axisymmetry around a Kerr black hole in ideal and resistive GRMHD frameworks. The initial conditions for the thin-disc are set following \cite{Dihingia-etal2021}. The accreting sub-Keplerian matter from the truncated accretion disc plays a crucial role in the flow dynamics and radiative properties around the black hole.

The simulations show turbulent mass loading features in the profiles of accretion rate, wind rate, and mass flux rates through the funnel due to the oscillation of inner accretion flow. They show quasi-periodic peaks in their profiles. We observe flow oscillating between MAD and non-MAD states rather than a fully developed magnetic arrested state. We also observed quasi-periodic magnetic eruption events following the accretion flow. 
During this process, the inflow and outflow develop shear and become unsteady. Such instabilities at the interface between inflow and outflow can facilitate the mixing of matter and may help in mass loading to the jet. In our simulations, we also observe the formation of a hot corona around the black hole due to shock transition in our simulations, which is intermittent in nature. The hot corona and the inner edge of the accretion disc mainly contribute to the observed high-energy emission. 

Our study supports the extended corona model during the evolutionary phase considered here. Also, the high energy emission modulates by orders of magnitude during the oscillation of the accretion flow. The modulations in the X-rays are often observed in terms of QPOs \citep[etc.]{Belloni-etal2011, Belloni-Motta2016, Ingram-Motta2019}. These are typical features of the hard-intermediate state (HIMS) of BH-XRBs. In the HIMS, intermediate photon index ($\alpha\sim1.8-2.4$, \cite[references therein]{Nandi-etal2012}) with occasional giant radio flares can be observed \citep[etc.]{Fender-etal2004,Fender-etal2009}. Some of the major findings from our extensive study are listed below,

\begin{itemize}
    \item[(1)] The qualitative features for all the ideal MHD models are similar. The reconnection events and the plasmoids formed due to active tearing mode instabilities play a crucial role in the dynamics of the accretion flow. They generate poloidal magnetic fields penetrating the equatorial plane and develop a strong magnetic tension force in the disc mid-plane. The resistance offered by this force halts the accretion flow momentarily. 
    With time, the ram pressure of the flow in the inner edge overcomes this force due to outward transport of angular momentum transport (due to MRI and disc-winds), leading to accretion again.
    This process continues throughout the simulation in the form of oscillations of the inner accretion flow. These oscillations are quasi-periodic in nature. The frequency of oscillations lie in the range of low-frequency QPOs (LFQPOs).
    The QPO frequency $(\nu_{\rm QPO, max})$ increases from $\sim13.5$ to $20$ Hz $(10M_{\odot}/M_{\rm BH})$ as the initial plasma-$\beta$ parameter increases from $\beta_{\rm tr}=100$ to $1000$. Similarly, the QPO frequency $(\nu_{\rm QPO, max})$ decreases from $\sim13.5$ to $7$ Hz $(10M_{\odot}/M_{\rm BH})$ as the truncation radius increases from $r_{\rm tr}=40$ to $60$.

    \item[(2)] In the resistive MHD models, we observe a qualitatively similar oscillating feature of the inner accretion flow as in the ideal MHD models. However, unlike the ideal MHD models, we do not observe the formation of plasmoids with the increase of resistivity. The turbulent features of the inner accretion flow flow are also subdued with the increase of magnetic resistivity. The KHI across the inflow and outflow boundaries is suppressed due to the increase in magnetic resistivity. 
    The frequency of oscillation increases with the increase of magnetic resistivity. Also, the radius of balancing magnetic tension force and the ram pressure $(r_{\rm max})$ decreases with the increase of magnetic resistivity.
    
    \item[(3)] We find that the high-energy emission comes mainly from the edge of the truncated accretion disc and the post-shock region. It suggests that the low angular momentum matter around the black hole plays a crucial role in the emission of hard X-rays. Subsequently, we find that the high-energy radiation modulates during the oscillation of the inner accretion flow. The emission within the energy range $h\nu\sim1-100keV (\nu\sim 10^{17}-10^{19}Hz)$ increases by two orders of magnitude during the oscillation. Similarly, we also find that the emission in the very high energy range $(h\nu>100KeV$ or $\nu >10^{19}Hz)$ increases around one order of magnitude during the oscillation. The 2D synthetic intensity maps at 1KeV show an edge brightened structure with most of the emission coming from the inner edge of the truncated accretion disc. We also observe two bright rings close to the black hole due to the presence of the hot post-shock corona with a typical post-shock proton temperature $T_p\gtrsim10^{11}$K.
\end{itemize}

\begin{figure}
\centering
\includegraphics[scale=0.45]{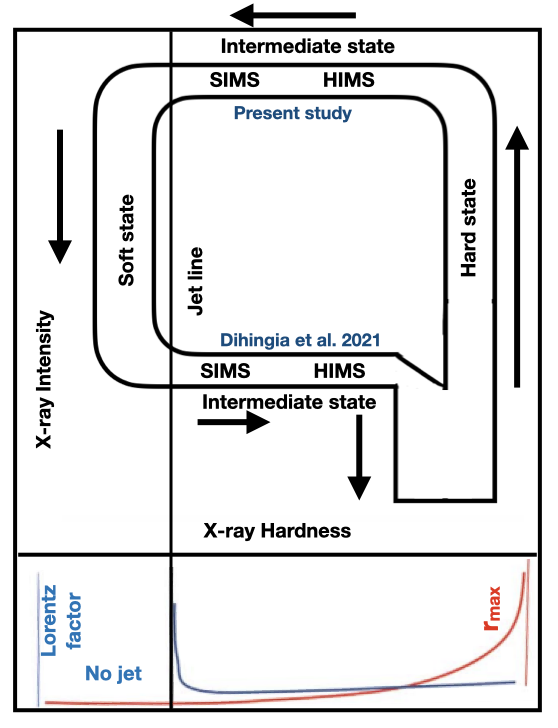}
\caption{Cartoon diagram for X-ray hardness-intensity diagram (HID) for a typical outburst in BH-XRBs. 
Different spectral states (hard, intermediate HIMS: hard-intermediate, SIMS: soft-intermediate, and hard state) are marked on the figure, and the arrows indicate the direction of the evolution with time.
The schematic evolution of the jet Lorentz factor and the disc inner radius $(r_{\rm max})$ is also shown at the lower panel. 
}
\label{fig-hid}
\end{figure}

The current study is useful in comprehending the 'Q' diagram of outburst BH-XRBs. 
A cartoon diagram of the same is shown in Fig.~\ref{fig-hid} for better understanding. 
The figure depicts various spectral states (hard, intermediate, and soft states), with arrows indicating the evolution's direction.  The intermediate state is further classified as HIMS (hard-intermediate state) and SIMS (soft-intermediate state).
\citep[see][]{Belloni2010}. The jet-line separates the intermediate and soft states, where the source exhibits peaks in radio emission as it approaches the jet-line and drastic changes in jet properties are observed along this line \citep{Fender-etal2009}. The schematic evolution of the jet Lorentz factor and the inner disc radius ($r_{\rm max}$) is shown in the lower panel of the cartoon diagram.
The inner edge of the high angular momentum material is responsible for the $r_{\rm max}$ (i.e., Keplerian matter). 
HIMS is associated with a stage in which high angular momentum material lies around the equatorial plane for a radius $r> r_{\rm max}$, while the material inside ($r < r_{\rm max}$) oscillates, resulting in periodic modulation in high energy emission. 
Giant radio flares are a common HIMS \citep{Fender-etal2009} signature. 
These HIMS characteristics closely resemble those of our current simulation models (as shown in Fig.~\ref{fig-hid}). 
Recently, the X-Ray transient MAXI J1803-298, which is hosted by a stellar-mass Kerr black hole X-ray binary, was observed in HIMS and found to have QPOs of the order of $\nu_{\rm QPO} \lesssim 10$\,Hz \citep{Chand-etal2022,Jana-etal2022}, which agrees closely with our simulation results. Earlier studies on low angular momentum transonic non-magnetized accretion flows also hinted at the possibility of such QPOs due to oscillations of shock waves \citep{Lee-etal2011, Lee-etal2016, Das-etal2014, Sukova-etal2017}.

Our study also shows that with time, the critical radius $r_{\rm max}$ reduces and eventually is expected to reach the event horizon/ISCO. 
Such a reduction in the radial extent of the high-angular momentum flow is even faster for cases with high magnetic resistivity. 
As the radial extent of the inner edge approaches the ISCO, the outburst transits into a soft state. 
In our earlier study \citep{Dihingia-etal2021}, the focus was to understand the outburst evolution from such a soft state, starting with a 
high angular momentum thin disc close to the black hole. 
We had observed a transition from a quasi-steady phase to an oscillatory phase (HIMS) (marked on the Fig. \ref{fig-hid}). 
Thus, both these studies cover a branch of the `Q' diagram whereby the outburst in an intermediate state evolves to a soft state (high-angular 
momentum disc close to the black hole) and further evolves to an intermediate state again \cite[e. g.,][]{Dunn-etal2010, Belloni2010, Belloni-Motta2016}. 
Ideally, to cover the complete path followed by the outburst, one would need to evolve the truncated accretion disc for a much longer
time duration with realistic boundary conditions at the outer boundary. 
Further, for a more consistent inference on radiative and timing properties, the thermodynamics of electrons (heating and cooling) 
and protons (heating and cooling) needs to be considered. 
Incorporation and application of such two-temperature models within the GRMHD simulations are currently under development (e.g. \cite{Ressler-etal2015,Ryan-etal2017,Sadowski-etal2017,Mizuno-etal2021,Dihingia-etal2022}). 
In the future, longer temporal evolution simulation studies with physically consistent thermodynamics would be crucial to understand the  complete `Q' diagram of an outburst in BH-XRBs.



\appendix
\section{MRI quality factor}
\begin{figure}
\centering
\includegraphics[scale=0.32]{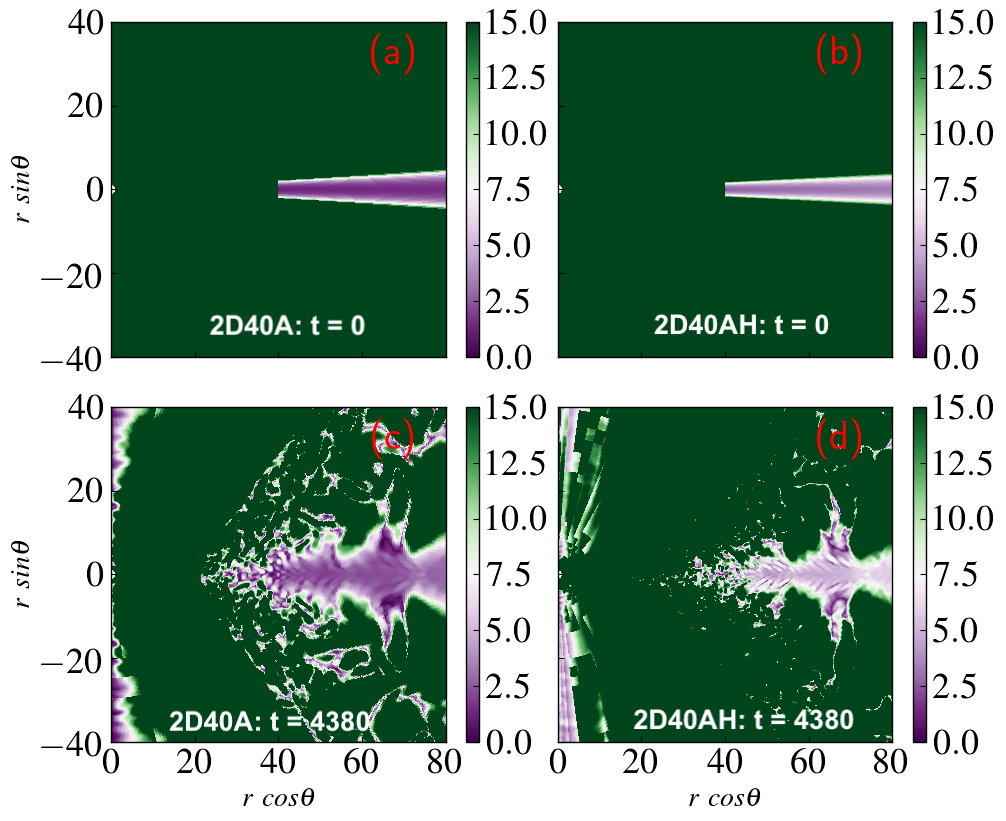}
\caption{Distribution of MRI quality $(Q_\theta)$ factor on the poloidal plane for model 2D40A at simulation times (a) $t=0$ and (c) $t=4380$, and for model 2D40AH at simulation times (b) $t=0$ and (d) $t=4380$.}
\label{fig-qtheta}
\end{figure}

To ensure that MRI is resolved for a given set of numerical resolutions, we calculate the MRI quality factor in terms of wavelength ($\lambda_\theta$) of the fastest growing MRI mode in the $\theta$ direction as $Q_\theta = \lambda_\theta/\Delta x_\theta$. The wavelength of the fastest growing MRI mode $\lambda_\theta$ is given in this case by
\begin{align}
\lambda_\theta = \frac{2\pi}{\sqrt{(\rho h + b^2)\Omega}}b^\mu e_\mu^{(\theta)},
\end{align}
and the grid resolution $\Delta x_\theta = \Delta x^\mu e_\mu^{(\theta)}$ (see \cite{Takahashi2008, Siegel-etal2013, Porth-etal2019, Nathanail-etal2020}, for details).
Typically, $Q_\theta \gtrsim 6$ is require to resolve this MRI mode (see \cite{Sano-etal2004}). In Fig. \ref{fig-qtheta}, we show the distribution of $Q_\theta$ for model 2D40A (resolution: $1024\times512$) at simulation times (a) $t=0$ and (c) $t=4380$, and for model 2D40AH (resolution: $2048\times1024$) at simulation times (d) $t=0$ and $t=4380$. The panels suggest that for our simulation runs, $Q_\theta$ sufficiently greater than $10$ in most of the numerical domain. For the low resolution model (2D40A, Fig. \ref{fig-qtheta}a and \ref{fig-qtheta}c), in the high density equatorial part is somewhat under-resolved $(Q_\theta\sim4-5)$. However, with the increase of resolution, in the high density equatorial part for is also resolved with $Q_\theta\sim6$ (Fig. \ref{fig-qtheta}b and \ref{fig-qtheta}d).

\section{3D low resolution comparison run}
\begin{figure*}
\centering
\includegraphics[scale=0.55]{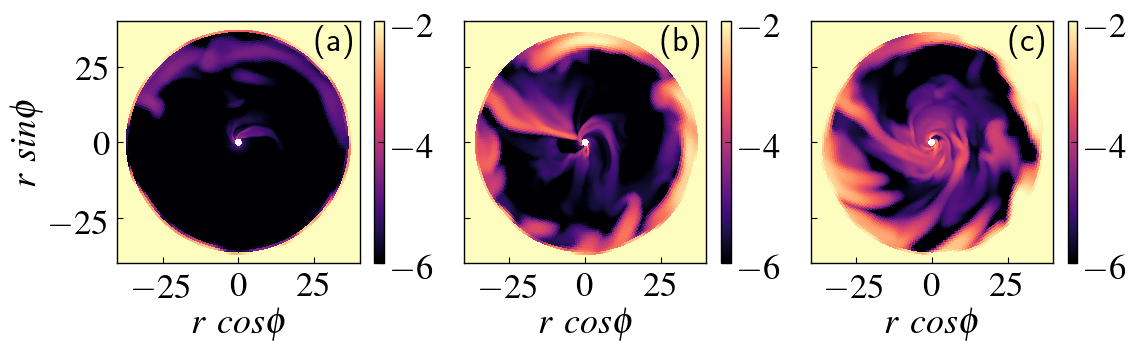}
\caption{Logarithmic density distribution on the equatorial plane for model 3D40 at time (a) $t=2500$, (b) $t=3500$, and (c) $t=4500$.}
\label{3D-density}
\end{figure*}

For completeness, we devised a 3D run of truncated accretion disc with a truncation radius $r_{\rm tr}=40$ and $\beta_{\rm tr}=100$ considering resolution $224\times96\times128$. All other initial conditions are the same as in the reference model. With time, non-asymmetric disc instabilities develop in the accretion disc \citep[e.g.,][]{Hawley1987, Savonije-etal1990}.
Depending on the active instability modes, matter starts to accrete from certain regions of the truncated accretion disc. As a result, the axisymmetric nature of the truncated accretion breaks, and accretion happens in terms of spiraling {\it fingers}. To study the evolution of the accretion {\it fingers} in detail, in panels of Fig. \ref{3D-density}, we show the distribution of logarithmic density ($\rho/\rho^{\rm tr}$) for model 3D40 at three different simulation times $t=2500, 3500$, and $4500$, in panels (a), (b), and (c), respectively. 

In Fig. \ref{3D-density}a, we observe a small active accretion {\it finger}. At this time, the accretion rate is expected to be at its minimum. With time, more matter accrete towards the black hole in terms of accretion {\it fingers} (see Fig. \ref{3D-density}b). In Fig. \ref{3D-density}c, we show a state where we observe abundant spiraling matter close to the black hole. Thus, the accretion {\it fingers} is highly dynamical. Also, it is interesting to note that the spiraling {\it fingers} are separated by a low-density region. The dynamics of both regions may interact and play a very interesting role in understanding physics around the black hole. 

\begin{figure}
\centering
\includegraphics[scale=0.4]{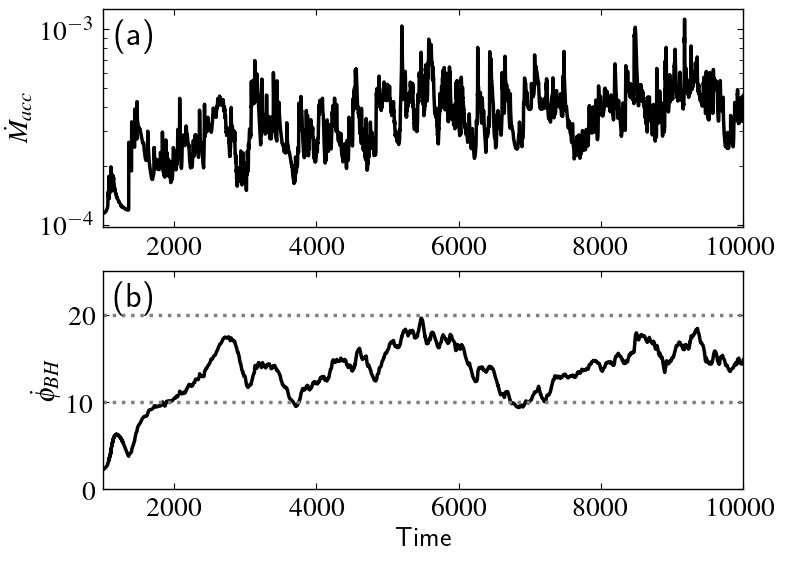}
\caption{Plot of accretion rate (a) $(\dot{M}_{\rm acc})$ and (b) magnetic flux $(\dot{\phi}_{\rm BH})$ for the 3D model, respectively. The horizontal lines in panel (b) correspond to $\dot{\phi}_{\rm BH}=10$ and $\dot{\phi}_{\rm BH}=20$. }
\label{3D-fluxes}
\end{figure}

The long-term temporal evolution of the 3D model is shown in Fig. \ref{3D-fluxes}, where we plot the accretion rate as a function of simulation time in panel Fig. \ref{3D-fluxes}a. 
In the figure, we observe that the accretion rate profile does not show clear sharp peaks in its profile as in the axisymmetric model. Rather, we observed broader peaks in the accretion rate profile. 
Similarly, in Fig. \ref{3D-fluxes}b, we show the plot of magnetic flux accumulated at the event horizon  $(\dot{\phi}_{\rm BH})$ for the 3D model as a function of simulation time. With the temporal evolution, magnetic flux started to accumulate around the event horizon, and $(\dot{\phi}_{\rm BH})$ increases. After a certain time $(t\sim2000)$, the inner accretion disc reaches a MAD configuration. After that ($t>2000$), the value of $\dot{\phi}_{\rm BH}$ remains in between within $\dot{\phi}_{\rm BH}\sim 10-20$, suggesting a fully developed MAD state in the flow \citep{Tchekhovskoy-etal2011}. 

In summary, despite the low resolution, we capture the salient features of the 3D truncated accretion disc. We observe that due to the presence of non-axisymmetric disc instabilities, the matter from the truncated accretion disc accretes in terms of spiral {\it fingers}. The dynamics of accretion {\it fingers} determines the dynamics of the accretion flow. Unlike the axisymmetric models, we do not observe periodic sharp peaks in the accretion rate profile in this case. Instead, we see broad peaks of uneven strength throughout the simulations. A recent high-resolution 3D study by \cite{Ripperda-etal2022} has shown the formation of magnetic flux bundles due to reconnection events. Thus, the observed broader peaks in our 3D runs may be due to the high numerical resistivity and absence of reconnection events. The peaks may become prominent with the increase in resolution.

\section*{Acknowledgements}
All simulations were performed on the Max Planck Gesellschaft (MPG) super-computing resources. 
We would like to thank the financial support from the Max Planck partner group award at Indian Institute Technology of Indore. IKD thank Swarnajayanti Fellowship from the Department of Science and Technology (DST/SJF/PSA-03/2016-17) for the financial support.

\section*{Data Availability}
The data underlying this article will be shared on reasonable request to the corresponding author.



\bibliographystyle{mnras}
\bibliography{references} 
%
\bsp	
\label{lastpage}
\end{document}